\documentclass[twocolumn]{aastex631}

\usepackage{courier}
\usepackage{multirow}
\usepackage[inline]{enumitem}
\usepackage{amsmath}
\usepackage{tabularx}
\usepackage{longtable}
\usepackage{graphicx}
\usepackage{xcolor}

\usepackage{enumitem}
\usepackage{amsmath}
\newcommand{\kms}{\ensuremath{\mathrm{km\,s^{-1}}}}
\newcommand{\gaia}{\textit{Gaia}}

\graphicspath{{./}{figures/}}

\shortauthors{Wu et al.}

\begin{document}

\title{Searching for Contact Binaries with LAMOST and TESS}

\author[0009-0007-1276-7144]{Ting Wu}
\affiliation{Xinjiang Astronomical Observatory, Chinese Academy of Sciences, Urumqi, Xinjiang 830011, People's Republic of China}
\affiliation{School of Astronomy and Space Science, University of Chinese Academy of Sciences, Beijing 100049, People's Republic of China}

\author[0000-0002-7420-6744]{Jin-Zhong Liu}
\affiliation{Xinjiang Astronomical Observatory, Chinese Academy of Sciences, Urumqi, Xinjiang 830011, People's Republic of China}
\affiliation{School of Astronomy and Space Science, University of Chinese Academy of Sciences, Beijing 100049, People's Republic of China}

\author[0000-0002-7135-6632]{Senyu Qi}
\affiliation{Department of Astronomy, Xiamen University, Xiamen, Fujian 361005, People's Republic of China}

\author[0000-0002-2419-6875]{Zhi-Xiang Zhang}
\affiliation{College of Physics and Information Engineering, Quanzhou Normal University, Quanzhou, Fujian 362000, People's Republic of China}

\author[0000-0001-5796-8010]{Hubiao Niu}
\affiliation{Xinjiang Astronomical Observatory, Chinese Academy of Sciences, Urumqi, Xinjiang 830011, People's Republic of China}

\author[0000-0003-1845-4900]{Ali Esamdin}
\affiliation{Xinjiang Astronomical Observatory, Chinese Academy of Sciences, Urumqi, Xinjiang 830011, People's Republic of China}
\affiliation{School of Astronomy and Space Science, University of Chinese Academy of Sciences, Beijing 100049, People's Republic of China}

\author[0000-0003-3137-1851]{Wei-Min Gu}
\affiliation{Department of Astronomy, Xiamen University, Xiamen, Fujian 361005, People's Republic of China}

\correspondingauthor{Jin-Zhong Liu; Wei-Min Gu}
\email{liujinzh@xao.ac.cn; guwm@xmu.edu.cn}

\begin{abstract}
Contact binaries (CBs) serve as fundamental laboratories for studying complex stellar interactions, including mass transfer, tidal effects, and angular momentum loss. In this work, we search for CB with high-precision light curves from the Transiting Exoplanet Survey Satellite (TESS) and large radial-velocity variation from the Large Sky Area Multi-Object Fiber Spectroscopic Telescope (LAMOST). We derive a sample of 1,281 CB candidates,  among which 266 are newly reported. Our sample with both high-precision photometry and medium-resolution spectra may provide new constraints on the physical scales, luminosity calibration, and population distribution of CBs, offering valuable insights into their evolutionary role within the stellar population.
\end{abstract}


\section{Introduction}\label{sec:intro}
Over half of the stars in the universe reside in binary or multiple systems, with contact binaries (CBs) representing a particularly fascinating subclass. These systems are characterized by both components filling their Roche lobes, leading to mass and energy exchange via a shared common envelope \citep{1959K,1979ApJ...231..502L}. While the common envelope redistributes energy and drives the two stellar surfaces toward a single effective temperature ($T_\mathrm{eff}$), the system as a whole is not in global thermal equilibrium. This persistent disequilibrium powers the thermal-relaxation oscillations that govern CB evolution \citep{1968ApJ...151.1123L,1976ApJ...205..217F}. The study of CBs thus offers crucial insights into mass transfer, angular momentum loss, and stellar merger processes.

CBs are further classified into two subtypes: A-type systems, in which the most massive star typically exhibits a higher $T_\mathrm{eff}$, and W-type systems, where the less massive star is hotter \citep{1970VA.....12..217B}. The distinct properties of these subtypes make CBs valuable probes for understanding stellar structure and evolution. They are closely linked to related phenomena, including fast-rotating FK Comae Berenices stars \citep{2007A&A...467..229P}, blue straggler stars \citep{1987ApJ...314..585K}, and luminous red novae like V1309 Sco \citep{2008IAUC.8972....1N,2011A&A...528A.114T}. 

Beyond their intrinsic importance, binary and contact systems play a central role in a wide range of astrophysical contexts: stellar physics \citep{2023A&ARv..31....1A}, gravitational wave astronomy \citep{2014MNRAS.445L..74M,2016PhRvL.116f1102A,2017PhRvL.119p1101A}, and cosmology \citep{2013Natur.495...76P,2014ApJ...780...59G}. Eclipsing binaries, in particular, allow precise determination of stellar parameters under well-defined geometric constraints and can even serve as standard candles under favorable conditions.

In recent years, large-scale surveys have revolutionized the classification and cataloging of variable stars, providing unprecedented data for the detecting and classifying CBs. For example, \cite{2018MNRAS.477.3145J} employed a random forest classifier, supplemented with visual verification, to identify and classify 66,179 bright variable stars from the All-Sky Automated Survey for Supernovae (ASAS-SN), including 2,427 EW-type systems. Similarly, \cite{2020ApJS..249...18C} used the Zwicky Transient Facility (ZTF) to classify 781,602 periodic variables in the northern sky, identifying approximately 369,707 as EW-type CBs. These studies demonstrate the feasibility of large-scale CB detection but also reveal key challenges—photometric blending, survey heterogeneity, and incomplete phase coverage—that limit classification precision and sample purity.

To overcome these limitations, we propose an integrated approach for the identification and characterization of CBs by combining multi-wavelength data. Specifically, we employ the period–radius relation as a diagnostic tool to identify candidate systems using high-precision photometry from the Transiting Exoplanet Survey Satellite (TESS) and stellar parameters from the Large Sky Area Multi-Object Fiber Spectroscopic Telescope (LAMOST). TESS provides unparalleled photometric precision and nearly all-sky coverage, while LAMOST contributes reliable spectroscopic parameters, enabling robust classification and physical characterization of CBs.

The paper is organized as follows. Section~\ref{sec:data and method} describes the data source process and methodology. Section~\ref{sec:results} presents the results of candidate identification and validation. Section~\ref{sec:discussion} discusses the findings and their implications. Finally, Section~\ref{sec:conclusions} summarizes the key findings and outlines potential directions for future research.

\section{Observational Data and Methodology}\label{sec:data and method}
The advent of long-duration, high-precision surveys in recent years has significantly advanced the identification and cataloging of CBs, yielding a substantial sample for systematic studies. In this research, we employed data from several prominent surveys to compile a representative sample of CB candidates.

\subsection{Photometric Observations}\label{sec:tess}
TESS is an all-sky survey mission led by the Massachusetts Institute of Technology and funded by NASA. It is designed to search for transiting exoplanets orbiting bright stars. Equipped with four red-sensitive wide-field cameras, TESS delivers a combined field of view of $24^{\circ}\times96^{\circ}$, enabling it to scan almost the entire sky in 27.4\,-day sectors \citep{2014SPIE.9143E..20R, 2015JATIS...1a4003R}. The TESS detectors are sensitive to wavelengths between 600 and 1000 nm, providing high-precision photometric data (typical 1-hour precision $\sim$ 70\,parts per million) broadly corresponding to the $I_\mathrm{C}$ band, making them well-suited for investigating eclipsing binaries and other types of stellar variability \citep{2015JATIS...1a4003R,2021A&A...649A..64B}.

\subsection{Astrometric Observations}
The \gaia\ mission, led by the European Space Agency (ESA), provides unparalleled precision in measuring stellar positions, parallaxes, and proper motions \citep{2016A&A...595A...2G}. \gaia\ Data Release 3 (\gaia\ DR3) provides improved astrometric and stellar parameter measurements, enabling detailed of stellar populations \citep{2023A&A...674A...1G}. For stars brighter than \(G=14\), \gaia\ achieves distance measurements with a median precision of 5\%, extinction estimates with uncertainty of $\sim$ 0.07-0.09\,mag, and $T_\mathrm{eff}$ determinations with uncertainty of $\sim$ 100-150\,K \citep{2023A&A...674A..27A}.

\gaia\ DR3 provides an extensive set of stellar parameters, including magnitudes in \(G\), \(G_{\mathrm{BP}}\), \(G_{\mathrm{RP}}\), as well as parallax, radii, $T_\mathrm{eff}$, luminosities, and proper motions \citep{2023A&A...674A...1G}. These high-precision measurements are critical for identifying and characterizing CBs, as they enable accurate determinations of distances and fundamental stellar properties. Additionally, derived parameters such as $T_\mathrm{eff}$ provide further constraints, aiding in the refinement of sample selection.

\subsection{Spectroscopic Observations} 
LAMOST \citep{2012RAA....12.1197C,2015RAA....15.1095L} is a spectroscopic facility that combines a large aperture with a wide field of view. Located at Xinglong Observatory (117$^\circ$34$'$30$''$E, 40$^\circ$23$'$36$''$N) at an elevation of 960 meters, LAMOST is equipped with 4,000 fibers and provides a $5^{\circ}$ field of view. In its medium-resolution mode, with a resolution power of R $\sim$ 7500, LAMOST provides enhanced spectral coverage in both the blue and red arms, spanning the wavelength ranges 4950 - 5350 \text{\AA} and 6300 - 6800 \text{\AA}, respectively \citep{2020arXiv200507210L,2012RAA....12..723Z}.

LAMOST provides reliable stellar parameters, such as $T_\mathrm{eff}$ and surface gravity (log \emph{g}), which play a crucial role in the accurate identification and characterization of binary systems. The LAMOST Stellar Parameter Pipeline achieves a precision of 110 K for $T_\mathrm{eff}$, and 0.19 dex for $\log g$, ensuring the reliability of these measurements \citep{2014w}. In addition to ensuring high-quality photometric and astrometric data, it is essential to develop a theoretical framework that provides a solid basis for the identification of CBs. 

\subsection{Theoretical Model for Binary System Periodicity}

The core of our search strategy for contact binaries relies on the physical constraint of the Roche lobe. For any star of a given mass and radius, there exists a minimum orbital period ($P_{\mathrm{orb}}^{\min}$) below which the star would exceed its Roche lobe. 
According to \citet{1983ApJ...268..368E}, the effective Roche-lobe radius ($R_{\mathrm{L,1}}$) is given by:
\begin{equation}\label{eq:eggleton}
    \frac{R_{\mathrm{L,1}}}{a} = \frac{0.49\,q^{2/3}}{0.6\,q^{2/3} + \ln\left(1+q^{1/3}\right)},
\end{equation}
where $a$ is the orbital semi-major axis and $q = M_1/M_2$ is the mass ratio. To derive an analytical selection criterion that is independent of the unknown companion mass ($M_2$), we adopt the simplified power-law approximation from \citet{1971ARA&A...9..183P}:
\begin{equation}\label{eq:paczynski}
    \frac{R_{\mathrm{L,1}}}{a} \approx 0.462 \left( \frac{M_{1}}{M_{1} + M_{2}} \right)^{1/3}.
\end{equation}

The orbital period $P_{\mathrm{orb}}$ is related to the semi-major axis $a$ and the total mass of the system via Kepler's Third Law:
\begin{equation}\label{eq:kepler}
    P_{\mathrm{orb}}^2 = \frac{4\pi^2 a^3}{G(M_1 + M_2)}.
\end{equation}
By assuming the primary star fills its Roche lobe ($R_1 = R_{\mathrm{L,1}}$) and substituting Equation \ref{eq:paczynski} into Equation \ref{eq:kepler}, the total mass term ($M_1 + M_2$) cancels out. This yields a lower limit for the $P_{\mathrm{orb}}$ that depends solely on the average density of the donor star:
\begin{equation}\label{eq:3}
    P_{\mathrm{orb}}^{\min} \approx 0.369 \left( \frac{M_1}{M_\odot} \right)^{-1/2} \left( \frac{R_1}{R_\odot} \right)^{3/2} \, \text{days}.
\end{equation}

If the observed photometric period of a target is shorter than this calculated $P_{\mathrm{orb}}^{\min}$, it strongly indicates that the single-star assumption is physically invalid and the system is likely an unresolved contact binary. 

We emphasize that this derivation serves as a first-order filter for our candidate selection. While Equation \ref{eq:paczynski} is traditionally optimized for $q < 1$, its application remains robust for our purposes. 

To improve the physical realism of this model, we incorporated the empirical mass-radius relation from \cite{1991Ap&SS.181..313D}, which distinguishes between scaling laws for stellar components depending on whether \(M_{1}\) is below or above \(1.66 \, M_\odot\). The relation is given as:
\begin{equation}
R_1 / R_\odot \approx \begin{cases} 
1.06 \, (M_{1} / M_\odot)^{0.945}, & M_{1} < 1.66 \, M_\odot, \\
1.33 \, ({M_{1} / M_\odot{}})^{0.555}, & M_{1} > 1.66 \, M_\odot.
\end{cases}
    \end{equation}
By combining this mass-radius relation with Kepler's third law, we derived a refined approximation for the theoretical lower limit of the $P_{\mathrm{orb}}$ as a function of stellar radius:
\begin{equation}
P_{\mathrm{orb}}^{\text{min}} \cong \begin{cases} 
0.381 \, (R_{1} / R_\odot)^{0.971}~\mathrm{days}, &\text{for } R_{1} < 1.711 \, R_\odot, \\
0.477 \, (R_{1} / R_\odot)^{0.599}~\mathrm{days}, & \text{for } R_{1} > 1.711 \, R_\odot.
\label{eq:p_r}
\end{cases}
\end{equation}
Systems with \(P_{\mathrm{orb}}\) $\leq$ \(P_{\mathrm{orb}}^{\text{min}}\) are classified as CB candidates, where the visible star fills its Roche lobe. Those with $P_{\mathrm{orb}} > P_{\mathrm{orb}}^{\text{min}}$ probably represent detached systems, where both components remain within their respective Roche lobes. To mitigate the potential bias from measurement uncertainties and intrinsic scatter near the boundary, we additionally display a conservative buffer curve, offset by $+0.1$\,dex in $\log P$ relative to the theoretical threshold, $P_{\mathrm{shifted}}(R)=10^{0.1}\,P_{\mathrm{orb}}^{\min}(R)$. 

\begin{figure*}
    \centering
    \includegraphics[width=2.0\columnwidth]{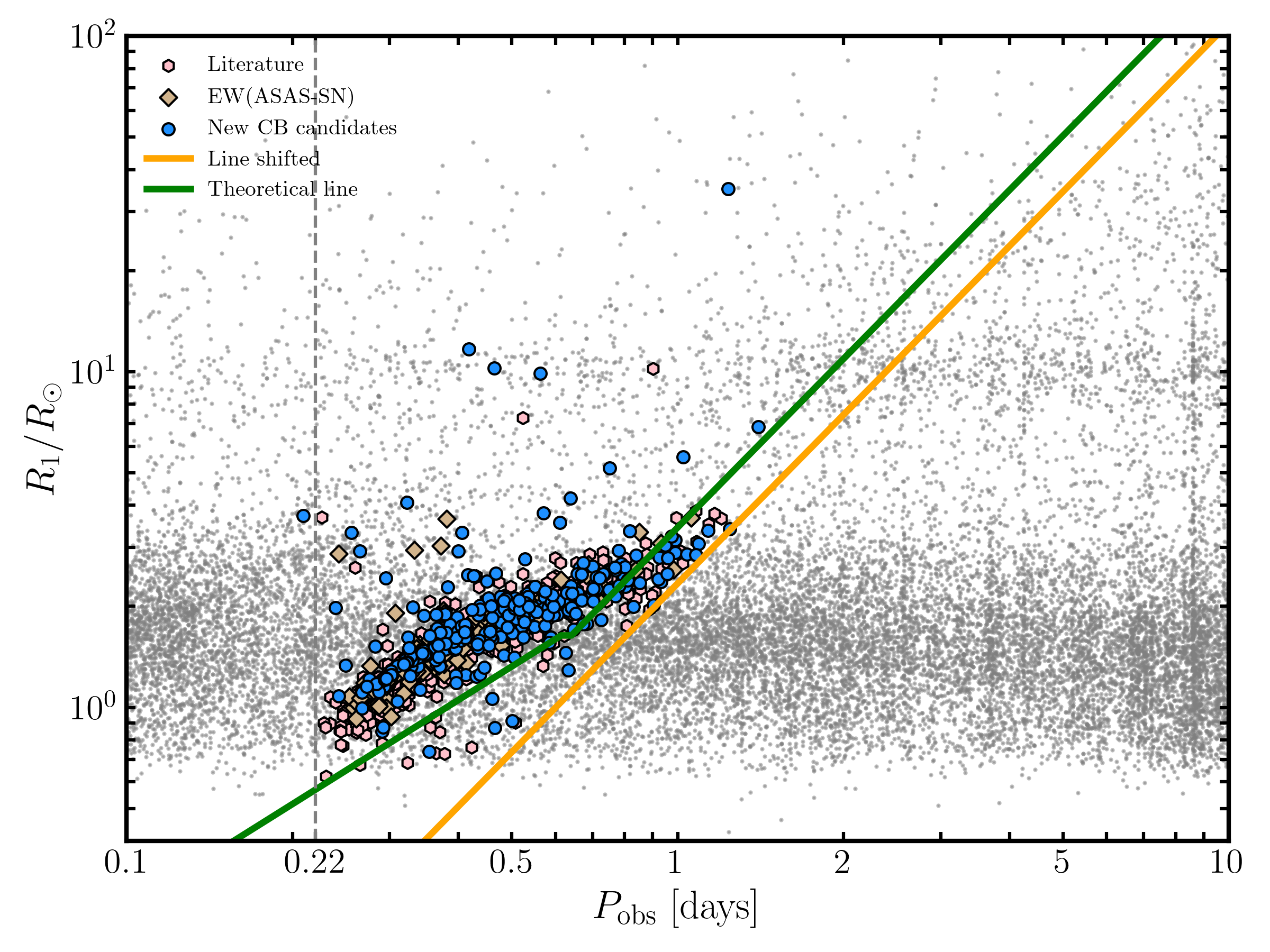}
    \caption{
    Period-radius distribution of CB candidates in our sample. \(P_\mathrm{obs}\) is derived from TESS light curve analysis, and \(R_{1}\) is obtained from \gaia\ DR3. The green solid curve represents the theoretical period-radius relation from Equation~\eqref{eq:p_r}. The orange curve is the same relation shifted right by $+0.1$\,dex in $\log P$. The vertical dashed line at \(P_\mathrm{obs}\) = 0.22 days represents the empirical minimum period for stable contact configurations \citep{2021MNRAS.503.3975P}. Systems located below this threshold are likely to be detached or merging binaries.}
    \label{fig:Period_radius}
\end{figure*}

\subsection{Sample Selection Criteria}\label{sec:sample}
We assembled our working sample by first cross-matching the full TESS source list with the LAMOST DR11 catalogue and then cross-matching the resulting set with \gaia\ DR3. The initial TESS-LAMOST match contains $\sim$\,2.2 million objects. To obtain a high-quality, well-characterized set of CB candidates, we applied the following sequential filters:
\subsubsection{{Data Quality Control}}
These steps focus on ensuring the reliability and quality of the data.
\begin{enumerate}[label={\arabic*.}]
\item Spectroscopic quality.
    Require S/N\,$>10$ in the LAMOST, leaving 1,060,295 sources.
\item Large radial-velocity variation.
    Select systems with $\Delta V_{\mathrm{R}}>30\ \mathrm{km\,s^{-1}}$ measured from the multi-epoch DR11 spectra, yielding 109,612 sources.
\item Precise astrometry and photometry.
    Cross-match with Gaia, retain bright, well-modeled stars by cutting at $G<14$ and $T_{\mathrm{eff}}<8000$ K, giving 52,260 objects. Period searches returned robust periods for 28,952 of these stars.
\end{enumerate}

\subsubsection{Physical Discrimination and Light Curve Validation}
These steps refine the sample based on their physical characteristics and the morphology of their light curves.
\begin{enumerate}[label={\arabic*.}]
\item Photometric periodicity.  
    Photometric data were obtained from TESS, and all available TESS light curves \citep{10.17909/t9-st5g-3177} were downloaded from the Mikulski Archive for Space Telescopes (MAST)\footnote[4]{\url{https://archive.stsci.edu}}. Data processing was performed using two independent pipelines: the TESS Science Processing Operations Center (SPOC; \citealt{2016SPIE.9913E..3EJ}) and the Quick Look Pipeline (QLP; \citealt{2020RNAAS...4..204H,2021RNAAS...5..234K}).
    Both the SPOC and QLP extractions were processed with the \texttt{lightkurve} toolkit. We search observed periods (\(P_{\mathrm{obs}}\)) in the range $0.01$–$27.4$ d with a Lomb–Scargle periodogram \citep{1976Ap&SS..39..447L,1982ApJ...263..835S} as implemented in \texttt{astropy.timeseries} module \citep{2018zndo...1227457P}. Uncertainties on the observed period were obtained by error propagation from the Lomb–Scargle peak’s half-maximum width.
\item CB candidates diagnostic. 
        For each source, we estimate $P_{\mathrm{orb}}^{\mathrm{min}}$, using equation~(\ref{eq:p_r}).
        Objects with $P_{\mathrm{obs}}\le P_{\mathrm{orb}}^{\mathrm{min}}$ are flagged as contact or near-contact systems, those slightly above the limit are retained as CB candidates. After automated filtering, 9,209 objects satisfy $P_{\mathrm{obs}}\le P_{\mathrm{shifted}}(R)$. 
\item Visual light curve inspection. All sources that passed the previous selection steps were visually inspected using their TESS light curves. Only those displaying the characteristic morphology of CBs, such as continuous, double-humped variations with two nearly equal minima, were retained. Systems showing detached features or dominated by noise or irregular variability were excluded.
\end{enumerate}

\section{Results}\label{sec:results}
\subsection{Period-Radius Distribution}

In this section, we present the period-radius distribution of the filtered sample. Figure~\ref{fig:Period_radius} shows the $P_{\mathrm{obs}}$ (in days) versus stellar radii (in solar radii) for the selected candidates. The stellar radii are taken from \citet{2022Gaia}, which are derived from Gaia parallaxes, extinction-corrected Gaia \(G\)-band magnitudes, and bolometric corrections \citet{2018A&A...616A...3R,2023A&A...674A..28F}. {Specifically, the radii are obtained from the GSP-Phot Aeneas best library using Gaia's BP/RP spectra, with the column name \texttt{radius\_gspphot} in \gaia\  DR3 and the radii are expressed in solar radii.}

The light-gray background points represent the full set of 28,952 systems with reliable TESS periods obtained during the filtering stage (Section~\ref{sec:sample}), providing a reference population for comparison. The green solid line in the diagram represents equation~\ref{eq:p_r}. The final catalog comprises 1,281 high-confidence CB candidates. We cross-matched our sample with several established CB catalogs: 667 overlapping systems from \citep{2020ApJS..249...18C}, 265 systems from the LAMOST catalog \citep{2020RAA....20..163Q}. We also cross-matched with ASAS-SN, obtaining 972 overlaps with our working list, to avoid double counting, if a system is in both the literature and ASAS-SN, it is coded as a literature CB. These confirmed CBs are shown as pink hexagons. ASAS-SN EW systems not present in the above literature lists are plotted as tan diamonds, while objects absent from all external catalogs and identified here for the first time are shown as blue circles. A total of 266 systems have no prior identification and are reported here as new candidates. A portion of the sample is presented in Table~\ref{tab: Table} for further reference.

The gray vertical dashed line in Figure~\ref{fig:Period_radius} represents this theoretical boundary. Systems below this limit are generally considered too compact to undergo Roche-lobe filling. Although previous studies suggest that the minimum period for CBs can reach below 0.22 days \citep{2011A&A...528A..90N, 2012MNRAS.425..950N, 2014ApJS..213....9D, 2013ApJ...764...62D}, this threshold is still a useful diagnostic tool for identifying candidates unlikely to be CBs.

We note that when this method is applied to binary systems, the single-star assumption may lead to overestimated radii for the binary components, which could affect their classification in the period-radius diagram. However, our goal is to develop a statistical diagnostic where the relative position of systems with respect to the period-radius boundary remains meaningful, despite these potential biases.

\subsection{Comparison of Photometric and Theoretical Minimum Orbital Periods}
\begin{figure*}
    \centering
    \includegraphics[width=2.0\columnwidth]{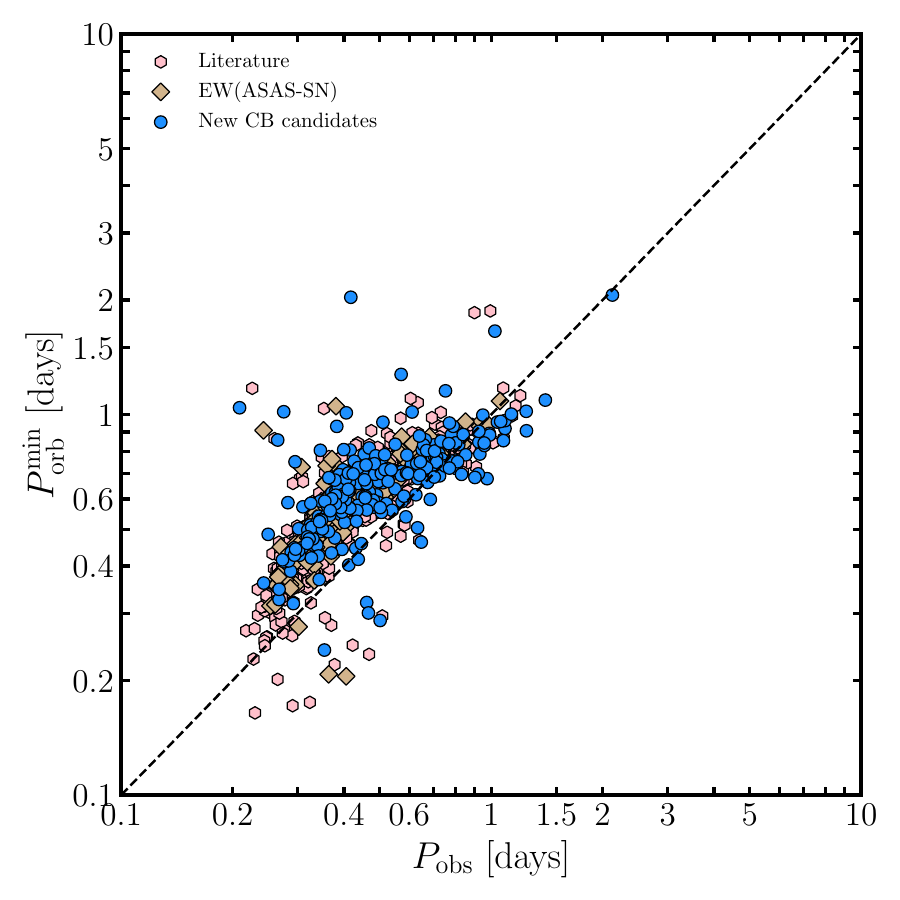}
    \caption{Comparison between \(P_\mathrm{obs}\) and \( P_{\mathrm{orb}}^{\mathrm{min}} \) for CB candidates. The gray dashed diagonal indicates equality between the two quantities. Symbols follow the same conventions as in Figure \ref{fig:Period_radius}.}
    \label{fig:Pobs_Porb}
\end{figure*}
To evaluate whether \(P_\mathrm{obs}\) of our candidates are consistent with Roche lobe filling, we recompute \(P_{\mathrm{orb}}^{\mathrm{min}}\) and compare it directly to \(P_\mathrm{obs}\), deriving distances from Gaia parallaxes, \(T_{\mathrm{eff}}\) and reddening from the Bayestar 3D dust map \citep{2019ApJ...887...93G}, bolometric corrections are from \cite{2018MNRAS.479L.102C}. These radii are then inserted into equation~\ref{eq:p_r} to compute 
\(P_{\mathrm{orb}}^{\min}\) for each object. Figure \ref{fig:Pobs_Porb} shows a systematic deviation between \(P_\mathrm{obs}\) and $P_{\mathrm{orb}}^{\mathrm{min}}$, with the majority of CB candidates exhibiting \(P_\mathrm{obs}\) $\leq$ $P_{\mathrm{orb}}^{\mathrm{min}}$. This trend is consistent with the expectation that these systems are in a gradual contact phase, where \(P_\mathrm{obs}\) reflects a dynamically evolving configuration approaching or reaching contact. 

This discrepancy can be attributed to several factors. The $ P_{\mathrm{orb}}^{\text{min}}$ is highly sensitive to stellar radii, which are subject to systematic uncertainties. {Unresolved binarity, referring to a binary system where the companion is not resolved,} tends to inflate the inferred radii by brightening the integrated photometry at fixed parallax, which in turn increases the estimated \(P_{\mathrm{orb}}^{\mathrm{min}}\). Extinction is uncertain along crowded or low-latitude sightlines, where the Bayestar 3D map may underpredict reddening, in our procedure the extinction is therefore capped using the Schlegel, Finkbeiner \& Davis all-sky map \citep{1998ApJ...500..525S}. In addition, imperfections in the bolometric-correction interpolation and in photometric $T_\mathrm{eff}$ tend to bias radii high when $T_\mathrm{eff}$ is underestimated or the extinction is overestimated. Finally, unresolved binarity brightens the spectral energy distribution at fixed parallax, inflating luminosities and therefore radii and $P_{\mathrm{orb}}^{\min}$.

The  \(P_\mathrm{obs}\) - $ P_{\mathrm{orb}}^{\text{min}}$ diagram is used to assess the consistency between the observed period and the  lower limit of the $P_{\mathrm{orb}}$. Despite the above deviations, the overall trend of the sample remains consistent with theoretical expectations. Future data releases from Gaia, with enhanced binary flags and precise dynamical mass estimates, are expected to significantly reduce uncertainties in stellar parameters and refine theoretical period predictions.

\subsection{Color-Magnitude Diagram}\label{sec: CMD}
\begin{figure*}
    \centering
    \includegraphics[clip,width=2.0\columnwidth,angle=0]{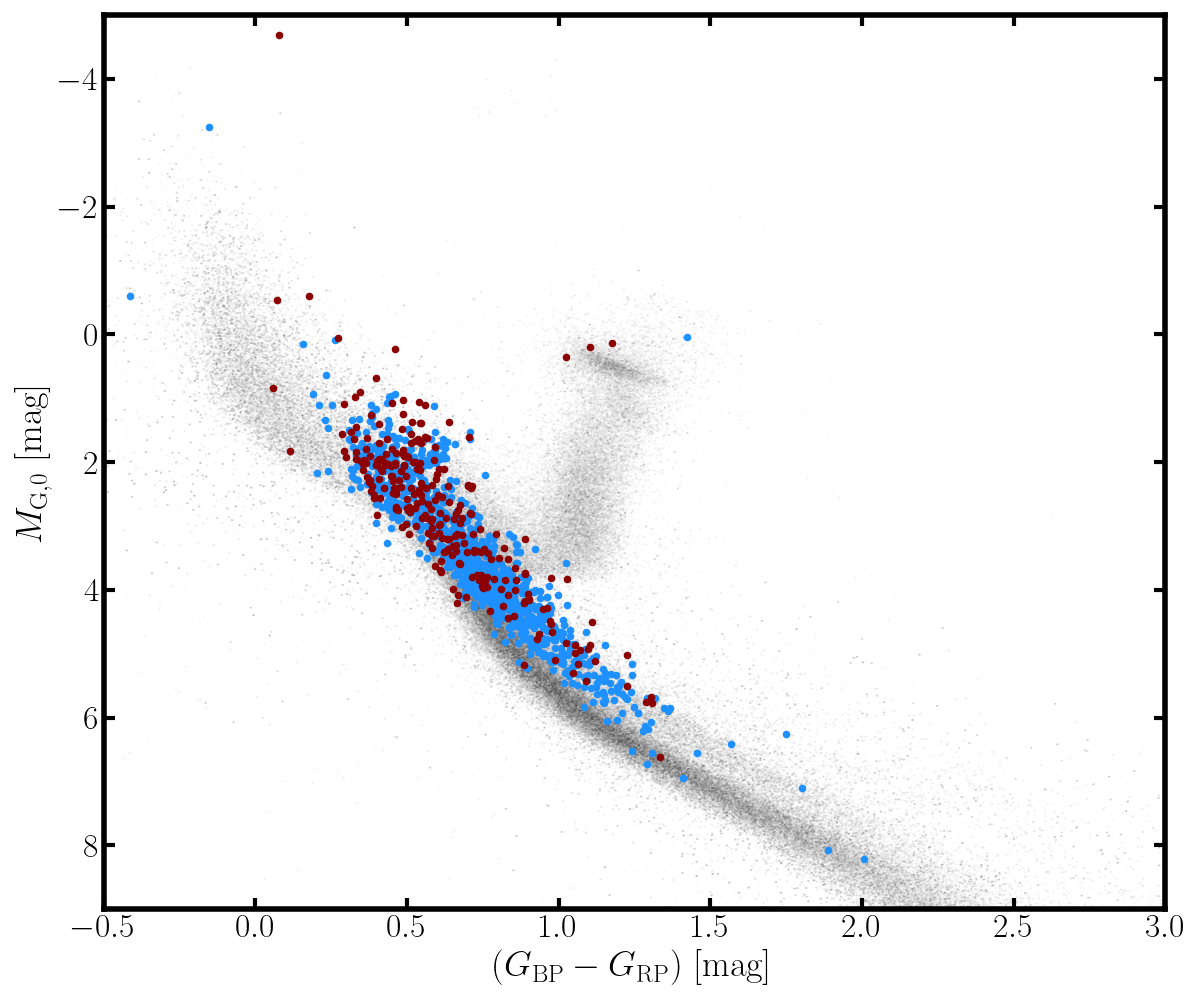}
    \caption{CMD of the samples. Gray points represent background stars, blue symbols denote literature-confirmed CBs, and red markers indicate newly identified CB candidates. The horizontal axis represents the \(G_{\mathrm{BP}}-G_{\mathrm{RP}}\) color index, while the vertical axis shows the extinction-corrected absolute magnitude \(M_\mathrm{G,0}\).}
    \label{fig:CMD}
\end{figure*}
The color-magnitude diagram (CMD) serves as a diagnostic tool for analyzing the photometric properties and evolutionary states of stellar populations. To construct the CMD, we combined photometric data from Gaia DR3 \citep{2023A&A...674A...1G} with spectroscopic parameters from APOGEE \citep{2022ApJS..259...35A}, selecting background field stars through an unbiased cross-matching procedure. Each star is characterized by its \(G_{\mathrm{BP}} - G_{\mathrm{RP}}\) color index and extinction-corrected absolute magnitude \(M_\mathrm{G}\), computed as:
\begin{equation}
    M_\mathrm{G} = m_\mathrm{G} + 5 \log \overline{\omega} - 10 - A_\mathrm{G},
\end{equation}
where \( m _\mathrm{G}\) denotes the apparent magnitude in \(G\)-band, \( \overline{\omega} \) represents the parallax from Gaia, and \( A_\mathrm{G}\) represents the \(G\)-band extinction. Extinction values were determined using the Bayestar19 3D dust map \citep{2019ApJ...887...93G}, accessed through the GALExtin service\footnote[6]{\url{http://www.galextin.org/}}, \( R_\mathrm{V} = A_\mathrm{V} / E(B-V) \), assuming a standard reddening law with \( R_\mathrm{V} = 3.1 \) \citep{2018ApJ...852...78W}. Band-specific extinction coefficients were applied as \(A_{\mathrm{G}_\mathrm{BP}}\) = 1.0139 \( A_\mathrm{V}\), \(A_{\mathrm{G}_\mathrm{RP}}\) = 0.5965\( A_\mathrm{V}\), and \(A_\mathrm{G}\) = 0.7889\(A_\mathrm{V}\). 

Figure~\ref{fig:CMD} illustrates the distribution of our sample in the CMD. The blue and red points correspond to previously identified CBs and newly identified CB candidates, respectively. As expected, binary candidates are generally located slightly above the main sequence. Additionally, a subset of outliers deviates significantly from the expected distribution. These objects may include evolved stars (e.g., subgiants or giants). Part of this systematic offset arises because our model assumes the primary star to be a main-sequence star, but uses Gaia-derived parameters for luminosity estimation. As a result, the luminosities of our sample are overestimated, causing a shift in the position of the stars in the CMD, with some points deviating slightly from the main sequence.

\begin{deluxetable*}{r r l r l r r r r l }

\tabletypesize{\scriptsize } 
    \tablecaption{Catalog of CB Candidates}
    \label{tab: Table}
    \tablehead{
    \colhead{R.A.} & \colhead{Decl.} & \colhead{TIC ID} 
    & \colhead{\(G\)} & \colhead{\(P_\mathrm{obs}\)} &\colhead{\(T_{\mathrm{eff,L}}\)} &\colhead{\(T_{\mathrm{eff,G}}\)}
    & \colhead{$\log g_\mathrm{,L}$}
    & \colhead{$\log g_\mathrm{,G}$}& \colhead{Catalog} \\
    \colhead{(deg)}  &\colhead{(deg)} &  &\colhead{(mag)} &\colhead{(day)} &\colhead{(K)} &\colhead{(K)}&\colhead{($\mathrm{cm\ s^{-2})}$
} &\colhead{($\mathrm{cm\ s^{-2}}$)
}& } 
    
\startdata
0.12684  &39.18524  &432482096 &13.65691 &0.28802(21)  &6770.7$\pm$28   &5654.9  &4.58$\pm$0.03   &4.36  &C20,ASAS-SN\\ 
0.40619  &54.53095  &346740777 &10.59059 &0.81716(175) &-               &6909.6  & -              &3.62  &N\\
0.49843  &-5.83590  &300887696 &13.09446 &0.37027(56)  &5880.8$\pm$42   &5337.2  &4.83$\pm$0.05   &4.19  &N\\
0.84341  &4.14644   &257503493 &13.16417  &0.30319(27)  &  -             &5676.3  &  -             &4.25  &C20,Q20,ASAS-SN\\ 
0.84913  &35.48414  &396393062 &13.13233 &0.39426(41)  &6100.5$\pm$27   &5811.5  &4.38$\pm$0.03   &4.08  &C20,ASAS-SN\\     
0.86662  &30.78778  &396382654 &13.18738 &0.38323(38)  &5852.8$\pm$41   &5841.9  &4.00$\pm$0.05   &4.16  &C20,ASAS-SN\\
1.05620  &14.11980  &395320045 &13.02195 &0.32304(34)  &5798.8$\pm$31   &5686.8  &4.72$\pm$0.04   &4.25  &C20,ASAS-SN \\       
1.06077  &31.25241  &396395188 &12.87841 &0.39805(41)  &7076.9$\pm$20   &6806.3  &4.13$\pm$0.02   &4.16  &C20,ASAS-SN \\
1.16012  &33.56835  &372130939 &13.73518 &0.67752(160) &6773.7$\pm$30   &6710.7  &4.17$\pm$0.04   &3.91  &C20,ASAS-SN\\  
1.55647  &36.44945  &194233735 &12.93081 &0.41291(59)  &6561.0$\pm$22   &6356.5  &4.28$\pm$0.02   &4.02  &J18,C20,ASAS-SN\\
1.76745  &33.90976  &354533594 &12.89599  &0.35396(44)  &6039.8$\pm$20   &5804.8  &4.49$\pm$0.02   &4.23  &C20,ASAS-SN\\  
2.05863  &33.58222  &283861424 &13.19827 &0.33499(39)  &4918.1$\pm$40   &5213.6  &2.42$\pm$0.04   &4.29  &C20,ASAS-SN \\     
2.23217  &30.70549  &283863167 &12.36230 &0.37816(37)  &6127.0$\pm$53   &5643.0  &4.07$\pm$0.07   &4.02  &C20,ASAS-SN\\ 
3.43713  &12.43376  &51943287  &12.25176 &0.44682(62)  &6462.0$\pm$21   &6246.3  &4.14$\pm$0.02   &4.04  &C20,ASAS-SN\\
4.25918  &35.45207  &365965666 &11.45217 &0.39102(40)  &5871.4$\pm$30   &6091.0  &4.11$\pm$0.03   &4.15  &A23\\
5.31158  &59.14600  &403261110 &12.19147  &0.40584(45)  &6661.2$\pm$22   &6146.9  &4.05$\pm$0.02   &3.57  &N\\
...      &...       &...       &...       &...          &...             &...     &...             &...   &... \\
350.99343 &9.78215  &354266695 &13.14216 &0.31828(33)  &5659.5$\pm$23   &5546.8  &4.18$\pm$0.02   &4.31  &ASAS-SN\\
351.20984 &8.67472  &354282100 &12.92370 &0.42866(59)  &5930.6$\pm$22   &5720.5  &3.79$\pm$0.02   &3.94  &C20\\
351.94180 &24.88248 &44383913  &12.60802 &0.73486(190) &7061.4$\pm$22   &5867.1  &4.20$\pm$0.03   &3.69  &C20,ASAS-SN\\
352.33170 &76.21464 &317597020 &13.31757  &0.36847(50)  &-               &5717.1  &-               &4.21  &N\\
352.73954 &61.98253 &270540858 &11.99151 &0.42363(48)  &-           	 &6174.6  &-               &4.01  &N\\
352.95823 &12.84942 &435249961 &12.15363 &0.49485(66)  &6431.2$\pm$63   &6333.6  &4.21$\pm$0.06    &3.89  &ASAS-SN\\
353.13569 &10.55532 &423329160 &12.54262 &0.34234(40)  &6203.6$\pm$20   &5943.8  &4.27$\pm$0.02    &4.04  &P03,C20,ASAS-SN\\
353.83723 &-0.36385 &313864690 &13.56890 &0.36333(43)  &5265.6$\pm$35   &5924.3  &2.95$\pm$0.04    &4.19  &C20,Q20,ASAS-SN\\
354.21023 &9.96369  &336627431 &13.44482 &0.28245(21)  &-               &5232.9  &-                &4.42  &C20,ASAS-SN\\
354.29456 &31.60314 &326448388 &13.26231 &0.37569(37)  &5258.4$\pm$70   &5810.5  &4.23$\pm$0.09    &4.35  &C20,Q20,ASAS-SN\\
354.84733 &15.97719 &336564632 &12.77808  &0.5207(100)  &7094.9$\pm$19   &6882.2  &4.24$\pm$0.02     &4.05  &ASAS-SN\\
354.92164 &14.47285 &336551417 &11.07990 &0.70157(37)  &-               &7667.0  &-                &4.01  &N\\
355.16188 &9.77410  &258617984 &13.01677 &0.36859(40)  &5441.8$\pm$37   &5709.3  &4.17$\pm$0.04     &4.14  &ASAS-SN\\
355.42246 &31.78257 &432341126 &10.71356 &0.36712(47)  &-               &6006.1  &-                &4.13  &D24,G25\\
355.75084 &15.10738 &434119021 &13.81257 &0.44826(54)  &6227.3$\pm$40   &5700.1  &3.76$\pm$0.05    &3.90  &C20,ASAS-SN\\
355.84164 &14.30928 &434120323 &13.83609  &0.31720(27)  &5720.5$\pm$38   &5755.8  &4.27$\pm$0.05    &4.31  &C20,ASAS-SN\\
355.99732 &0.45373  &293303799 &13.69525 &0.38337(60)  &-	        	 &6155.4  &-	              &4.24  &C20,Q20,ASAS-SN\\
356.01424 &38.80609 &352649814 &12.30201 &0.67912(120) &8012.9$\pm$16   &7807.4  &4.06$\pm$0.02    &3.82  &C20,ASAS-SN\\
356.08344 &35.07925 &288148191 &13.78217 &0.28890(1362)&5289.5$\pm$52   &4875.6  &4.89$\pm$0.07    &4.40  &C20,ASAS-SN\\
356.34017 &34.13910 &288226367 &11.12554  &0.35486(33)  &5521.3$\pm$69   &5586.3  &4.86$\pm$0.06    &4.05  &D24,G25,ASAS-SN\\
356.86710 &62.44145 &272928428 &12.93955  &0.37472(37)  &5676.5$\pm$45   &5651.6  &4.07$\pm$0.05    &4.21  &C20,Q20,ASAS-SN\\
356.99413 &39.92035 &455729273 &13.54347  &0.36709(35)  &-        	     &6418.4  &-                &3.92  &ASAS-SN\\
357.11998 &10.91321 &408410236 &12.34990 &0.47738(74)  &-               &7025.2  & -               &4.16  &C20,ASAS-SN\\
357.48262 &58.42262 &346304142 &12.58353 &0.65292(148) &6206.4$\pm$79   &6218.4  &3.80$\pm$0.09    &3.94  &N\\  
358.08012 &31.21641 &125281008 &13.01062 &0.36154(45)  &5433.9$\pm$37   &5435.7  &3.86$\pm$0.04    &4.22  &C20\\
358.09070 &57.46306 &65172690  &12.63776 &0.56005(82)  &7104.1$\pm$60   &7634.1  &4.34$\pm$0.08    &3.95  &C20,Q20\\
358.34618 &2.51839  &422752192 &13.67282  &0.28661(40)  &-               &4761.3  & -               &4.47  &ASAS-SN\\
359.14369 &31.66652 &155204110 &12.29164 &0.47654(59)  &5961.6$\pm$29   &6195.2  &3.57$\pm$0.03    &3.97  &C20,ASAS-SN\\
359.24094 &33.11908 &155224799 &13.72630 &0.36708(47)  &5626.2$\pm$40   &5474.1  &4.23 $\pm$0.05   &4.19  &N\\
359.25539 &16.59597 &2053351336&15.02682 &0.72180(210) &6879.7$\pm$19   &-       &4.30$\pm$0.02    &-     &N\\
359.53797 &37.17439 &291835887 &11.67236 &0.95735(239) &6572.0$\pm$48  &6469.4  &4.13$\pm$0.04    &3.77  &N\\
359.73274 &33.69263 &155260065 &13.73827 &0.34137(41)  &5307.6$\pm$35   &5509.9  &3.85$\pm$0.04    &4.23  &C20,ASAS-SN\\
359.94403 &37.61754 &455787295 &13.07182 &0.63850(110) &7027.0$\pm$30   &6819.3  &4.07$\pm$0.03    &3.96  &ASAS-SN\\
\\  \enddata
\begin{flushleft}
Note. Column 1: R.A. (J2000); Column 2: decl. (J2000); Column 3: TESS Input Catalog identifier; Column 4: Gaia \(G\)-band magnitudes; Column 5: periods and errors derived from TESS photometry; Column 6: \(T_{\mathrm{eff,L}}\) and errors from LAMOST; Column 7: \(T_{\mathrm{eff,G}}\) from Gaia; Column 8: $\log g_\mathrm{,L}$ and errors from LAMOST; Column 9: $\log g_\mathrm{,G}$ from Gaia; Column 10: catalog identifier for the candidates.
Literature references in Column 10: (P03) \cite{2003CoSka..33...38P}, (G06) \cite{2006AJ....131..621G}, (R07) \cite{2007AJ....134.2353R}, (Y13) \cite{2013MNRASY}, (K17) \cite{2017RMxAA..53..235K}, (Z19) \cite{2019ApJS..244...43Z}, (C20) \cite{2020ApJS..249...18C}, (Q20) \cite{2020RAA....20..163Q}, (L21) \cite{2021ApJS..254...10L}, (L22) \cite{2022xL}, (K22) \cite{2022ApJS..262...12K}, (P23) \cite{2023Pmass}, (L23) \cite{2023PASP..135e4201L}, (D24) \cite{2024AJ....167..192D}, (G25) \cite{2025ApJS..276...57G}, (ASAS-SN) \citep{2018MNRAS.477.3145J,2019MNRAS.486.1907J,2019MNRAS.485..961J,2023MNRAS.519.5271C}, (N) Newly identified CB candidates from this work.
\medskip
(This table is available in its entirety in machine-readable form.)
\end{flushleft}
\end{deluxetable*}

\section{Discussion} \label{sec:discussion}

\subsection{Cross-Correlation Function (CCF) Analysis}\label{sec:ccf}
We identified a sample of CB candidates through a multidimensional analysis of their physical properties, including period-radius relations and CMD derived from Gaia photometric data. To validate the binary nature of photometrically selected CB candidates, we performed a CCF analysis using LAMOST medium-resolution spectra, a robust method to determine radial velocities and resolve binary components.

The CCF is defined as:
\begin{equation}
    \mathrm{CCF}(\nu|F,G) = \frac{\mathrm{Cov}(F,G(\nu))}{\sqrt{\mathrm{Var}(F)\mathrm{Var}(G(\nu))}},
\end{equation}
where \( F \) is the normalized observed spectrum, and \( G(\nu) \) represents a PHOENIX synthetic template \citep{2013A&A...553A...6H}, Doppler-shifted by a relative velocity \( \nu \). The CCF quantifies the similarity between the observed and synthetic spectra, with values ranging from $-1$ (perfect anticorrelation) to $+1$ (perfect correlation). For binary systems, the CCF profile typically exhibits multiple peaks corresponding to the radial velocities of the stellar components.

Our analysis spans a velocity range of $\pm 500~\mathrm{km\,s^{-1}}$ with a step of $1~\mathrm{km\,s^{-1}}$. To optimize computational efficiency, synthetic templates are convolved to match the resolution of LAMOST medium-resolution spectra and resampled onto the instrument wavelength grid. The CCF is calculated using the $\texttt{laspec}$ pipeline from \cite{2021ApJS..256...14Z}, which includes continuum normalization and sky subtraction.

Synthetic spectra were generated using the PHOENIX model grid \citep{1997ApJ...483..390H,2013A&A...553A...6H}, with atmospheric parameters (\(T_{\mathrm{eff}}\), log \emph{g}) derived from Gaia DR3 and LAMOST DR7. These templates are specifically tailored to the LAMOST blue-arm spectral coverage, which includes strong metallicity-sensitive lines (e.g. Fe I at 4900 \text{\AA}, Mg I triplet) critical for radial velocity measurements in FGK-type stars \citep{2020RAA....20...51Z}.

\begin{figure*}
\centering \includegraphics[clip,width=2.0\columnwidth,angle=0]{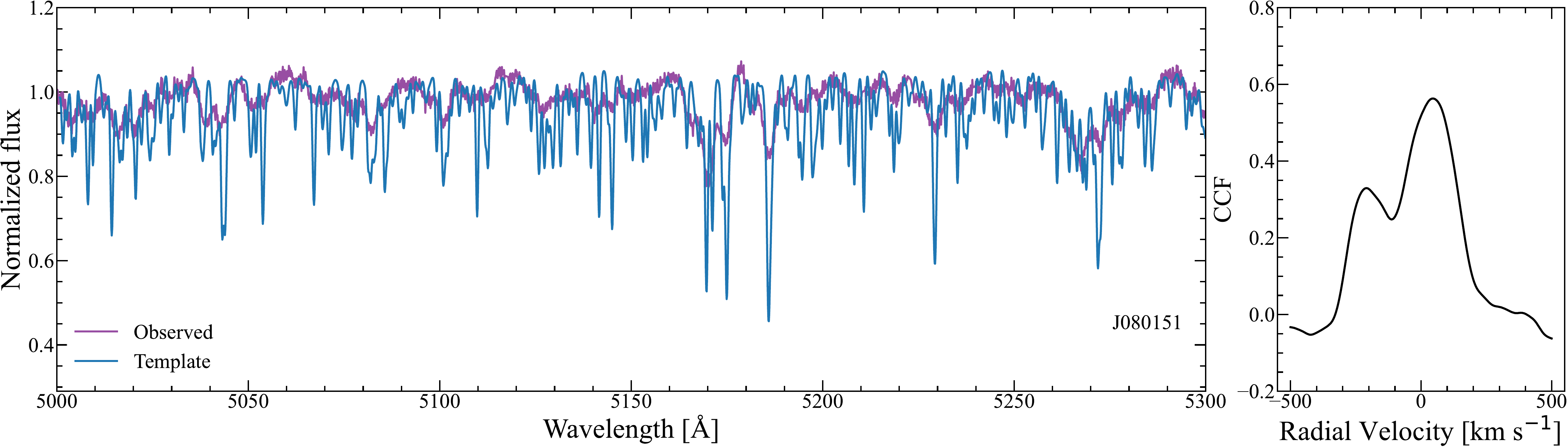}
\caption{CCF analysis of CB candidate J080151. The normalized medium-resolution LAMOST spectrum is overplotted with the best-fitting PHOENIX synthetic template, marked with distinct colors. The resulting CCF profile in the right panel highlights two distinct peaks, each corresponding to a binary component.}
\label{fig:ccf}
\end{figure*}

Figure~\ref{fig:ccf} demonstrates the CCF analysis for a representative candidate, J080151.52+413235.4 (hereafter referred to as J080151). The double-peaked profile unambiguously confirms its binary nature, with each peak corresponding to one of the binary components. Applying this method to the newly identified sample of 266 candidates, we confirm 246 systems as spectroscopic binaries based on resolved double lines in multi-epoch medium-resolution spectra. For the remaining objects without resolved double peaks, inspection of their spectroscopic phases reveals insufficient quadrature coverage. Therefore, their single-peaked CCFs are more likely due to sampling and resolution limitations rather than indicating a non-binary nature.

\subsection{Radial Velocity Determination}

We attempted to measure the RV curves of the binary system from spectra in order to constrain the mass ratio, component masses, and other physical properties. For short-period contact binaries, however, extracting radial velocities via the CCF is challenging. Due to the rapid rotation of both visible components, the absorption lines are significantly broadened, resulting in very wide CCF profiles. In cases where the relative velocity separation between the two stars is small, the peaks corresponding to the individual components merge into a single broad feature, making it impossible to distinguish a double peak \citep{Zucker1994}. Consequently, reliable radial velocities can only be obtained near quadrature phases ($\sim$0.25 and $\sim$0.75).

To extend the phase coverage of RV measurements, we adopted a direct spectral fitting approach, modeling each observed spectrum with a two-component synthetic template 
\citep{Gonz2006,elbadry2018}. The free parameters of the fit include the $T_\mathrm{eff}$ and metallicities of the two stars, the line broadening (including both rotational and instrumental effects, approximated here with a Gaussian kernel), the RVs of the two components, and the flux ratio. As spectral templates, we employed the MARCS library \citep{Gustafsson2008}, with interpolation performed using the stellarSpecModel tool \citep{zhang2025}.

\begin{figure*}
\centering \includegraphics[width=1.0\textwidth]{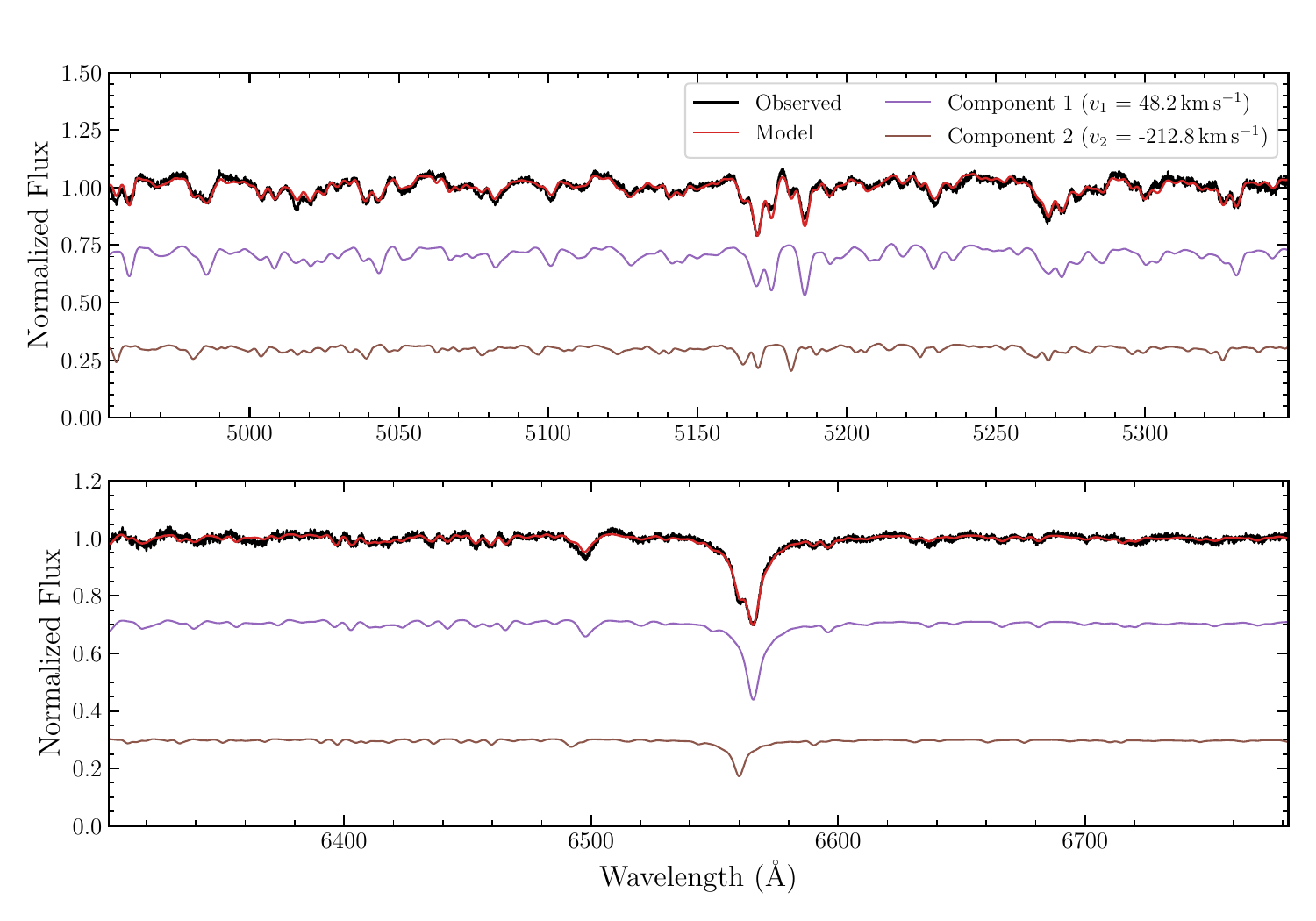}
\caption{Spectral decomposition of the CB candidate J080151. The top and bottom panels display the LAMOST medium-resolution spectra in the blue and red arms, respectively. In each panel: The black curve represents the observed spectrum. The red curve is the composite best-fitting model, combining contributions from both stellar components. 
The purple and brown curves correspond to the individual synthetic spectra of component 1 and component 2, respectively.
The close agreement between the observed and model spectra demonstrates the reliability of the two-component fitting approach.}
\label{fig:decomp}
\end{figure*}

We found that when the RV separation between the components is small, strong degeneracies arise between line broadening and RV, significantly increase the scatter of the RV curves. To mitigate this effect, we adopted a two-step fitting strategy. In the first step, all parameters were left free to vary, yielding a set of preliminary solutions. In the second step, the broadening, $T_\mathrm{eff}$, and metallicities were fixed to the median values from the first step, and the RVs were refitted. The RV curves obtained from this second iteration were adopted as our final results.

As illustrated in Figure~\ref{fig:decomp}, the LAMOST medium-resolution spectra of J080151 are well reproduced by a two-component composite model fitted simultaneously to both the blue and red arms. The purple and brown curves show the individual synthetic spectra of components 1 and 2, respectively, indicating two different RVs. This example demonstrates that our spectral decomposition, together with the two-step fit, recovers both components' velocities even when the CCF peaks are blended, thereby enabling the construction of RV curves for subsequent Keplerian modeling.

The derived RVs were then fitted to a Keplerian orbital model to describe the motion of the binary components. 
Considering the circular nature of the orbits in contact binaries, we employed a simple sinusoidal function to model the RV curves of both stars:
\begin{equation}
            \begin{cases}
            v_1(t)=K_1\bigl[\sin(\omega t+\phi)\bigr]+\gamma\\
            v_2(t)=-K_2\bigl[\sin(\omega t+\phi)\bigr]+\gamma
        \end{cases}
\end{equation}
where $v_1(t)$ and $v_2(t)$ are the RVs of the two components as functions of time, $K_1$ and $K_2$ are the velocity semi-amplitudes, $\omega = 2\pi/P_\mathrm{orb}$ is the angular frequency (with $P_\mathrm{orb}$), $\phi$ is the phase offset, and $\gamma$ is the systemic velocity of the binary. The sign difference reflects the anti-phase motion of the two stars in a circular orbit. We used the $\texttt{curve$\_$fit}$ function from the $\texttt{scipy.optimize}$ package to perform a least-squares fitting of the RV data to this model, allowing us to derive the orbital parameters, including the semi-amplitudes $K_1$ and $K_2$, and systemic velocity $\gamma$.

\begin{figure} 
\centering  
\includegraphics[clip,width=1.0\columnwidth,angle=0]{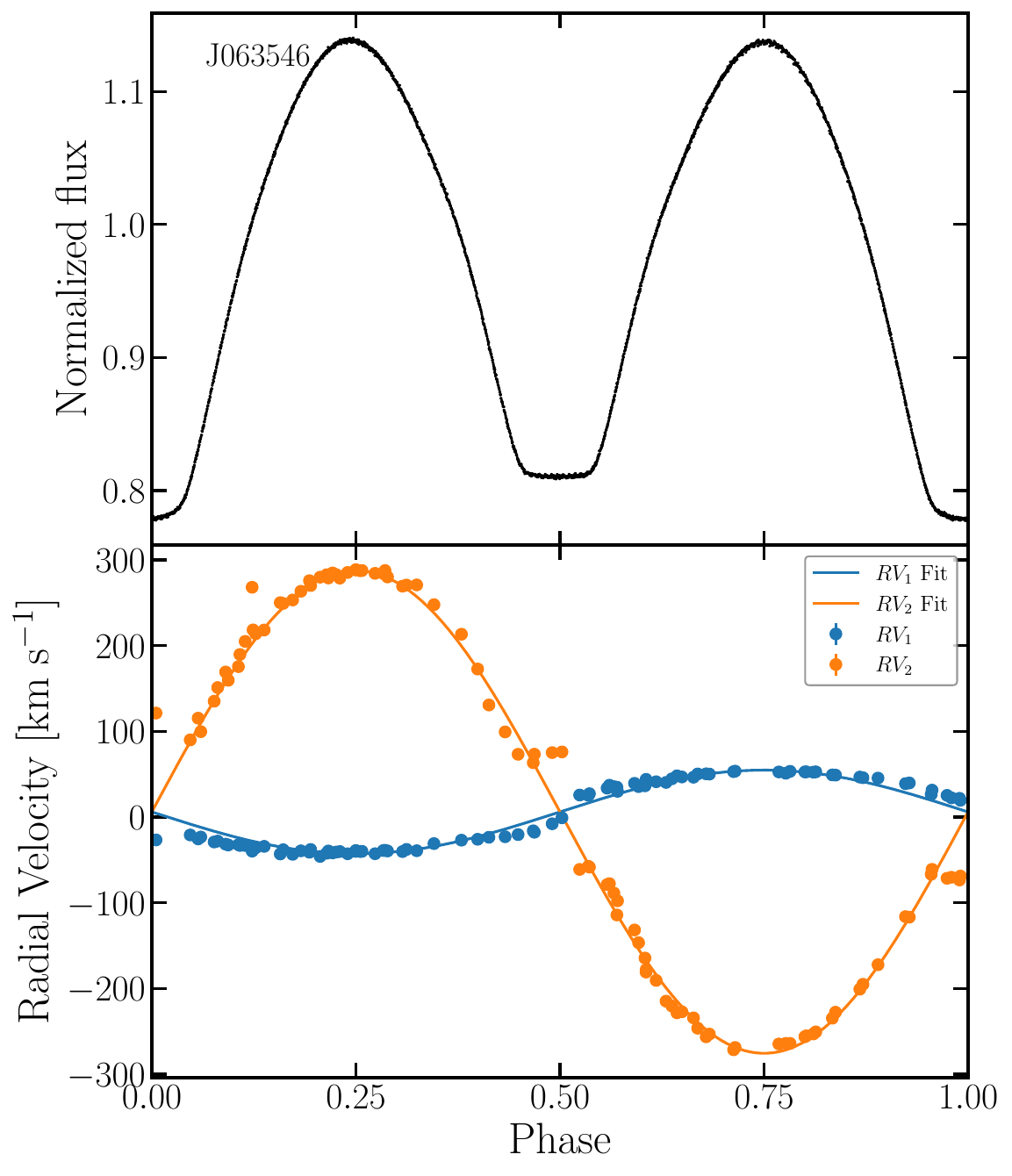}
\includegraphics[clip,width=1.0\columnwidth,angle=0]{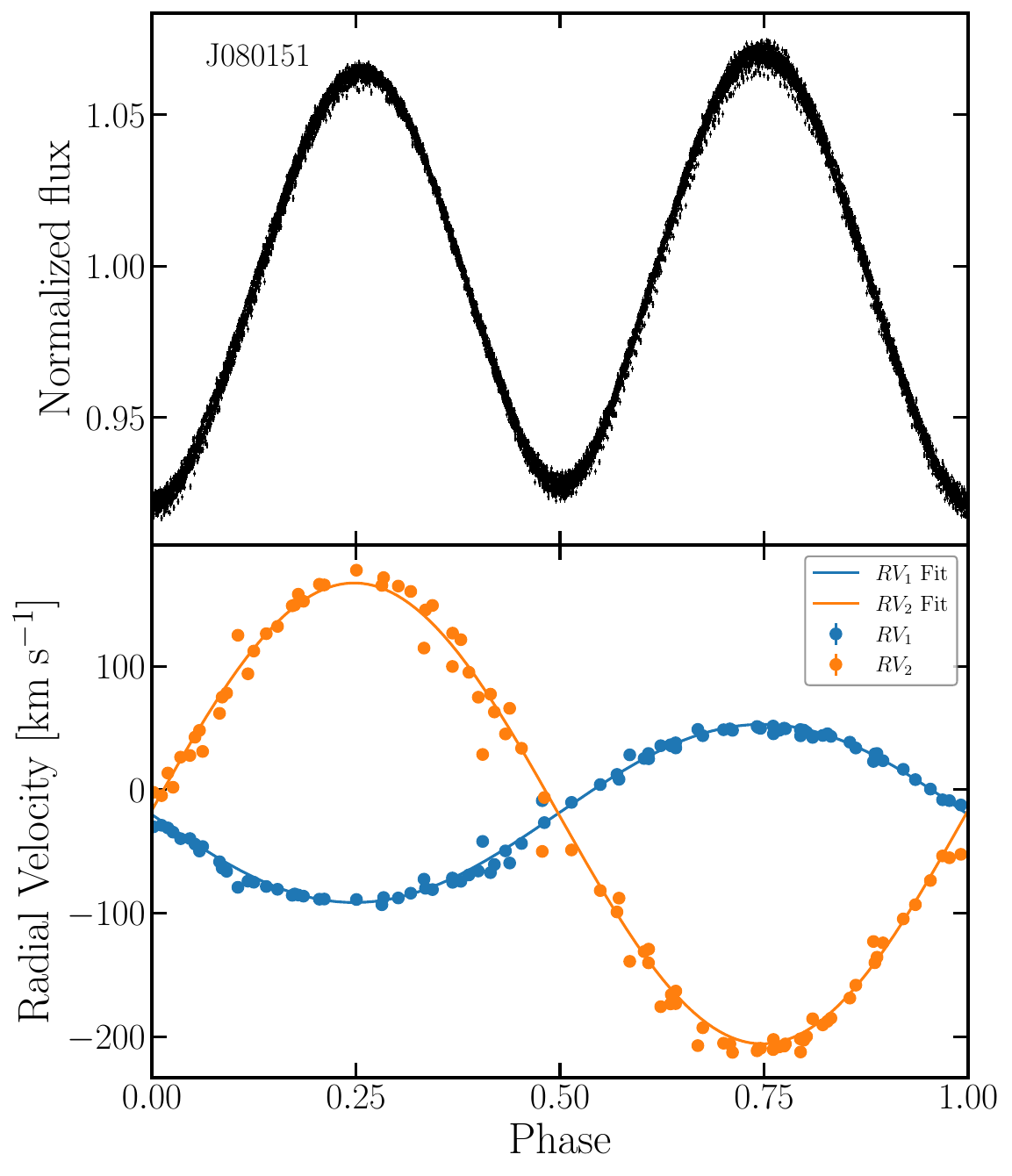}
\caption{Photometric and spectroscopic modeling of the binary systems. Top: TESS light curves of J063546 and J080151, showing clear flux variations consistent with eclipsing or ellipsoidal variability. Bottom: radial velocity curves for the two binary systems, overplotted with the best-fit Keplerian models for both components. Observed radial velocity points are shown with symbols, while solid lines represent the fitted curves.} 
\label{fig:lc_rv}
\end{figure}

Figure~\ref{fig:lc_rv} illustrates the results of this fitting process. The top panel presents the TESS light curves for the candidate stars J063546 and J080151, while the bottom panel shows the best-fit RV curves for the binary systems. 
For J063546, the RV amplitudes are $K_1 = 47.5\,\kms$ and $K_2 = 286.0 \,\kms$, indicating an extreme mass ratio of $q = 0.17$.
For J080151, the RV amplitudes are found to be $K_1 = 74.2 \,\kms$ and $K_2 = 191.3 \,\kms$. 
In addition to the two targets, {164 additional samples in our dataset were successfully folded to produce RV curves.} The RV curves of these objects are presented in Appendix \ref{appendix:rv_fits}, and the corresponding measurements, including the RV semi-amplitudes, mass ratios $q$, and mass functions, are listed in Appendix \ref{appendix:K}.

\section{Conclusions}\label{sec:conclusions}
This study advances in our understanding of CBs, primarily through the expansion of the CB sample, which is crucial for refining theoretical models of binary star evolution. By integrating data from TESS photometry, LAMOST spectroscopy, and Gaia astrometry, we derived a robust sample of 1,281 CB candidates, among which 266 sources are newly reported. This extensive sample provides new observational constraints on the physical properties and evolutionary paths of CBs. The methodology used to identify these systems has proven effective and reliable, establishing a solid foundation for future studies in this field.

Our analysis reveals key trends, including a well-defined period-radius relation and a photometric period versus theoretical minimum period diagram, reinforcing the validity of our sample selection criteria. The positions of these systems in the CMD suggest their close evolutionary proximity, with most binaries located slightly above the main sequence due to the combined light from both components. These results highlight the roles of mass transfer and angular momentum loss in shaping their long-term evolution.

Radial velocity measurements from LAMOST DR11 spectroscopy confirmed the binary nature of 246 of the 266 candidates. While the majority of these systems exhibit double-lined profiles in CCFs, a small fraction of stars may not display clear double peaks. For the remaining objects, the available spectroscopy does not adequately sample quadrature; rotational broadening and small velocity separations result in single-peaked CCFs. Although their TESS light curves show contact-like morphology, spectroscopic confirmation requires additional epochs near quadrature, and we therefore retain them as photometric CB candidates pending follow-up.

The CBs in our sample exhibit characteristics indicative of Roche lobe filling or near-contact configurations, consistent with ongoing mass transfer and angular momentum loss. The observed period-radius distribution reinforces these findings and provides important new constraints on theoretical models of CB evolution. In conclusion, this study represents a significant step forward in the characterization of CBs, both through the expansion of the sample size and the refinement of the selection criteria. The improved observational constraints provided by this study will be invaluable for refining stellar population models and advancing our knowledge of binary star evolution.

\section*{Acknowledgements}
We thank the anonymous referee for constructive suggestions that improved the paper.
This work was supported by the National Key R\&D Program of China for the Intergovernmental Scientific and Technological Innovation Cooperation Project under No. 2022YFE0126200, the National Key R\&D Program of China under grant 2021YFA1600401, and the National Natural Science Foundation of China under grants 12433007 and 12221003, the Tianshan Talent Training Program through the grant 2023TSYCCX0101, and the Central Guidance for Local Science and Technology Development Fund under No. ZYYD2025QY27.

We acknowledge the use of publicly available data from TESS, accessed through MAST, with funding provided by NASA's Science Mission Directorate. We also utilized data from LAMOST, a National Major Scientific Project operated by the Chinese Academy of Sciences, funded by the National Development and Reform Commission, and managed by the National Astronomical Observatories of the Chinese Academy of Sciences. Additionally, data from Gaia, processed by the Gaia Data Processing and Analysis Consortium (DPAC, \url{https://www.cosmos.esa.int/web/gaia/dpac/consortium}) with the support of national institutions involved in the Gaia Multilateral Agreement, were incorporated into this study. This research also benefited from resources provided by NASA's Astrophysics Data System (ADS), the SIMBAD database (operated at CDS, Strasbourg, France), and the VizieR catalog access tool (CDS, Strasbourg, France).

\section*{Data Availability}

The spectroscopic data used in this study are available from LAMOST DR11 at \url{https://www.lamost.org/dr11/v1.0/}. The astrometric and photometric data used for cross-matching are available from Gaia DR3 at \url{https://www.cosmos.esa.int/gaia}. Photometric data from the TESS mission are publicly accessible at \url{https://tess.mit.edu/observations/} and through the Mikulski Archive for Space Telescopes (MAST) at \url{https://archive.stsci.edu/tess/}. Time-domain data from the ASAS-SN are available at \url{https://asas-sn.osu.edu/}. 

\vspace{5mm}

\software{
    \texttt{astropy} \citep{2013A&A...558A..33A, 2018AJ....156..123A, 2022ApJ...935..167A}, 
    \texttt{astroquery} \citep{2019AJ....157...98G},
    \texttt{Lightkurve} \citep{2018ascl.soft12013L},
    \texttt{Laspec} \citep{2021ApJS..256...14Z},
    \texttt{Matplotlib} \citep{Hunter:2007},
    \texttt{NumPy} \citep{harris2020array},
    \texttt{Pandas} \citep{mckinney-proc-scipy-2010},
    \texttt{SciPy} \citep{2020NatMe..17..261V}
    }

\appendix

\setcounter{figure}{0}
\renewcommand{\thefigure}{A.\arabic{figure}}
\setcounter{table}{0}
\renewcommand{\thetable}{A.\arabic{table}}
\section{The RV fitting results}
\label{appendix:rv_fits}

The following figures show the RV curve fitting results. {The complete figure set (164 images) is available in the online journal.}    

\figsetstart
\figsetnum{A1}
\figsettitle{THE RV FITTING RESULTS}

\figsetgrpstart
\figsetgrpnum{A1.131}
\figsetgrptitle{RV curves of J154835.02+432844.9}
\figsetplot{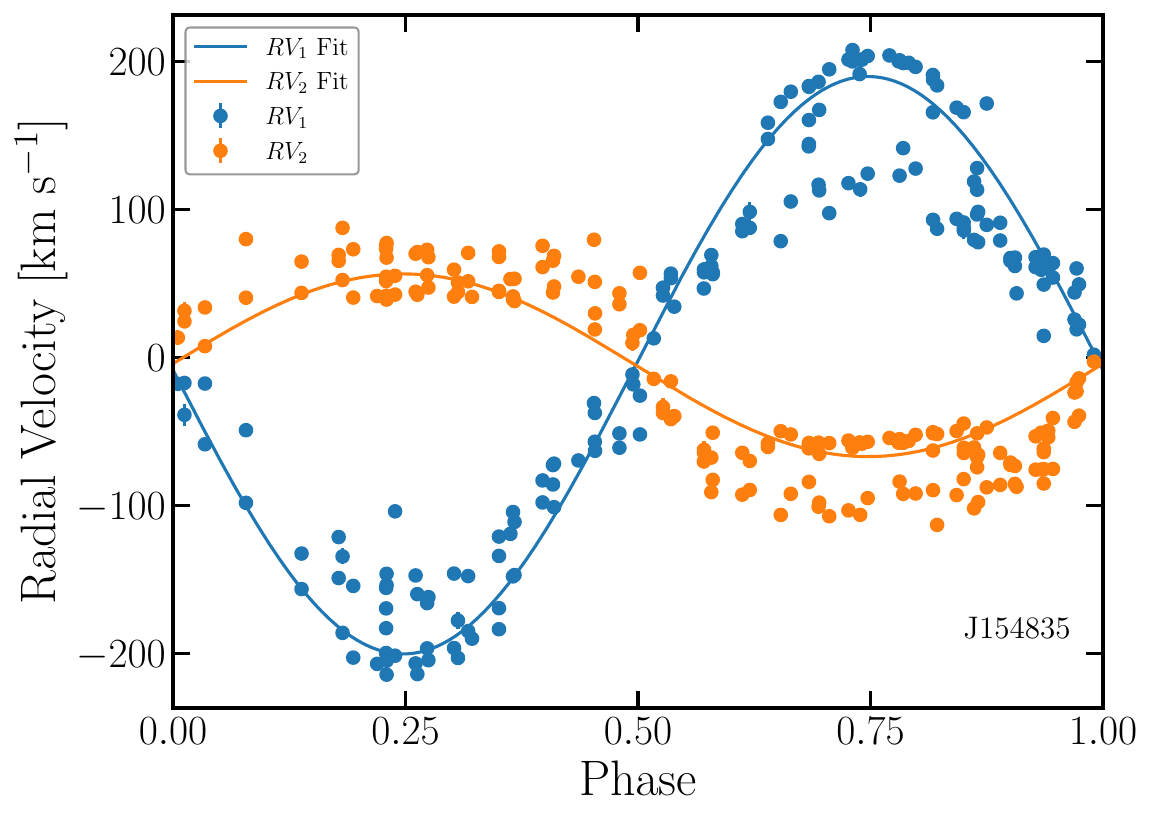}
\figsetgrpnote{RV curves for binary systems }
\figsetgrpend

\figsetend

\begin{figure}
\figurenum{A1}
\plotone{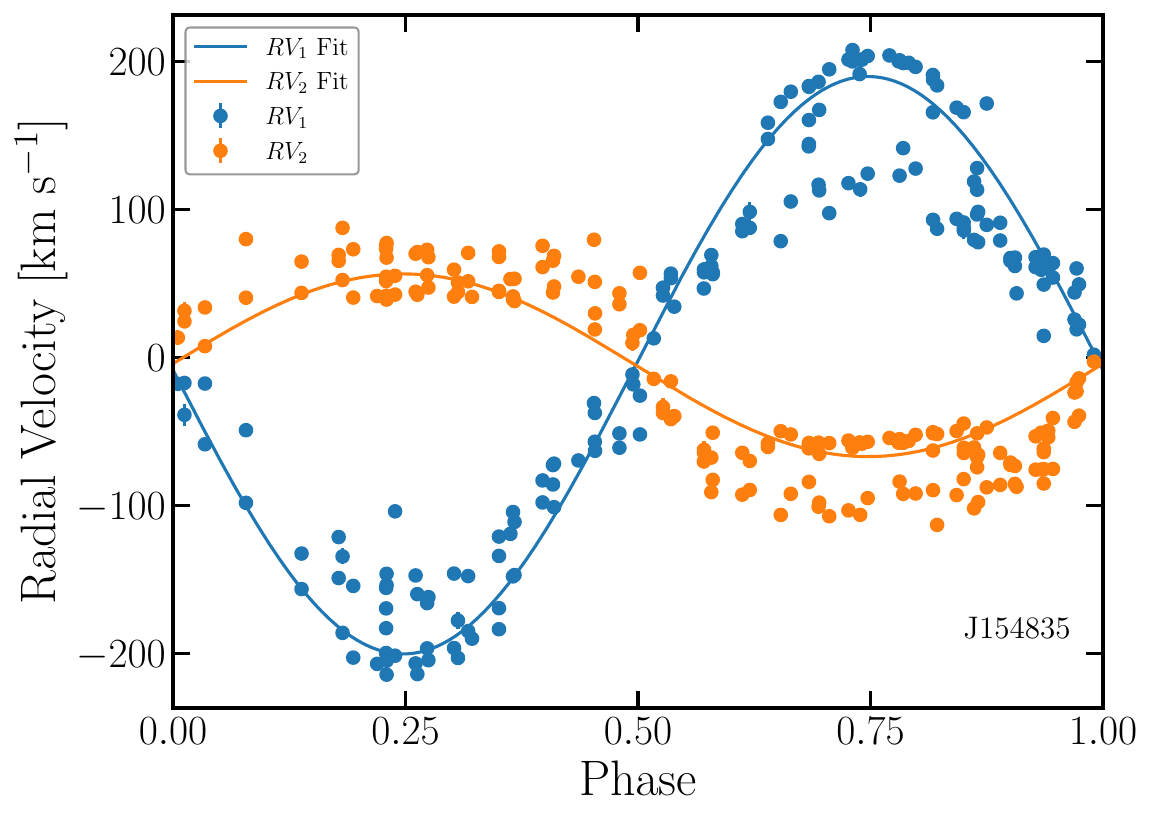}
\caption{RV curves for binary systems, with observed data points represented by symbols and solid lines depicting the fitted curves.}
\end{figure}

\clearpage
\setcounter{table}{0}
\renewcommand{\thetable}{B1}
\section{RV Amplitudes for Remaining Binary Systems}
\label{appendix:K}

The following table lists the RV semi-amplitudes and other parameters of the 166 binary systems.

\begin{table}[htbp]
    \centering
    \begin{tabular}{crrlrrlll}
        \hline
        Objname & R.A. & Decl.  & \(P_\mathrm{obs}\) &\( K1 \)&\( K_2\)& \( q \) & $f(M_1)$&\( f(M_2) \)\\
                & (deg)     &  (deg)   & (day)   &(\kms)    &(\kms)    &      &($M_\odot$) &($M_\odot$)\\
        \hline
        J004616 &  11.57060 &   8.17632 & 0.33115(34)  &  42.3$\pm$2.3   &  98.5$\pm$11.7  & 2.33(30) & 0.003(0)  & 0.033(12)\\
        J005425	&  13.60671 &   8.19478 & 0.26293(21)  &  99.0$\pm$13.9  & 110.8$\pm$16.1  & 1.12(23) & 0.026(11) & 0.037(16)\\	
        J005851 &  14.71657 &   3.06608 & 0.34249(46)  &  90.9$\pm$2.0   & 229.8$\pm$2.7   & 2.53(6)  & 0.027(2)  & 0.431(15)\\	
        J005903 &  14.76650 &   5.85914 & 0.27264(23)  &  97.9$\pm$6.7   & 218.1$\pm$8.0   & 2.23(17) & 0.027(5)  & 0.293(32)\\	
        J010209 &  15.53818 &  59.92862 & 1.16473(430) &  38.4$\pm$5.9   &  50.7$\pm$5.0   & 1.32(24) & 0.007(3)  & 0.016(5)\\
        J010658	&  16.74190 &   9.90470 & 0.37627(47)  & 103.7$\pm$20.0  & 214.6$\pm$12.3  & 2.07(42) & 0.043(25) & 0.385(66)\\
        J010802	&  17.01140 &   2.02380 & 0.41468(51)  &  94.4$\pm$5.1   & 204.3$\pm$5.4   & 2.16(13) & 0.036(6)  & 0.366(29)\\
        J010918	&  17.32895 &  10.40036 & 0.70475(146) &  52.6$\pm$3.5   &  71.7$\pm$4.3   & 1.36(12) & 0.011(2)  & 0.027(5)\\
        J010936	&  17.40120 &   4.18987	& 0.37567(42)  &  87.2$\pm$11.5	 &  94.5$\pm$4.8   & 1.08(15) & 0.026(10) & 0.033(5)\\
        J011207	&  18.03253 &   2.28614	& 0.35331(37)  &  58.2$\pm$6.9	 & 180.2$\pm$5.6   & 3.10(38) & 0.007(3)  &	0.214(20)\\
        J011836	&  19.65141 &  43.44236	& 0.33599(31)  &  84.1$\pm$3.6	 & 151.5$\pm$6.2   & 1.80(11) & 0.021(3)  & 0.121(15)\\
        J013006	&  22.52555 &  63.58255	& 0.46206(90)  &  93.1$\pm$9.2	 & 177.1$\pm$9.3   & 1.90(21) & 0.039(11) &	0.266(42)\\
        J013409	&  23.53947 &  45.95035	& 0.33456(30)  & 140.2$\pm$13.7	 & 145.3$\pm$13.0  & 1.04(14) &	0.096(28) &	0.106(29)\\
        J013437	&  23.65448 &  45.16077	& 0.43587(69)  & 137.0$\pm$26.1	 & 168.9$\pm$17.9  & 1.23(27) & 0.116(66) &	0.218(69)\\
        J014557	&  26.49032 &  37.12208	& 0.50676(93)  &  44.4$\pm$2.8	 & 54.5$\pm$3.0	   & 1.23(10) &	0.005(1)  &	0.008(1)\\
        J023824	&  39.60327 &  32.12823	& 0.66327(119) &  63.9$\pm$9.5	 & 65.9$\pm$10.1   & 1.03(22) &	0.018(8)  &	0.020(9)\\
        J024004	&  40.02006 &  35.16712	& 0.36353(36)  &  55.1$\pm$7.7	 & 114.1$\pm$6.4   & 2.07(31) &	0.006(3)  &	0.056(9)\\
        J025236	&  43.15110	&  54.59686	& 0.30511(25)  & 100.0$\pm$5.7	 & 217.8$\pm$6.6   & 2.18(14) & 0.032(5)  & 0.327(30)\\
        J031519	&  48.83280 &  55.26079	& 0.37679(38)  &  56.1$\pm$4.3	 & 162.1$\pm$4.7   & 2.89(24) & 0.007(2)  & 0.166(14)\\
        ......  &  ......   &  ......   & ......       &  ......         & ......          & ......   & ......    & ......   \\
        J221535 & 333.89970 &  28.66868	& 0.31957(37)  &  79.8$\pm$9.5	 & 105.5$\pm$5.8   & 1.32(17) & 0.017(6)  & 0.039(6)\\
        J221621	& 334.08808	&  29.55199	& 0.35027(33)  &  56.1$\pm$3.3	 & 213.4$\pm$3.2   & 3.80(23) &	0.006(1)  & 0.353(16)\\
        J221736	& 334.40179	&  28.11428	& 0.38647(54)  &  94.2$\pm$11.7	 &  96.8$\pm$13.5  & 1.03(19) & 0.033(12) & 0.036(15)\\
        J221915	& 334.81298	&  30.10086	& 0.47463(80)  &  55.7$\pm$7.6	 &  83.0$\pm$9.3   & 1.49(26) & 0.008(3)  &	0.028(9)\\
        J222153	& 335.47242	&  28.04632	& 0.31187(26)  &  96.2$\pm$14.4	 & 151.1$\pm$26.4  & 1.57(36) & 0.029(13) & 0.111(58)\\
        J222211	& 335.54677	&  29.59775	& 0.34592(43)  &  79.2$\pm$4.4	 &  96.7$\pm$8.3   & 1.22(12) & 0.018(3)  &	0.032(8)\\
        J222228	& 335.62038	&  29.36984	& 0.27798(21)  &  87.0$\pm$14.4	 & 165.4$\pm$7.5   & 1.90(33) & 0.019(9)  &	0.130(18)\\
        J222417	& 336.07191	&  28.66399	& 0.37402(38)  &  72.2$\pm$10.5	 &  80.5$\pm$10.5  & 1.11(22) & 0.015(6)  &	0.020(8)\\
        J225015	& 342.56433	&  34.78751	& 0.53621(103) &  30.0$\pm$11.3  &  42.8$\pm$5.9   & 1.43(57) &	0.002(2)  & 0.004(2)\\
        J225434	& 343.64501	&  34.69563	& 1.13360(459) &  65.8$\pm$3.1	 &  83.0$\pm$2.3   & 1.26(7)  &	0.033(5)  &	0.067(6)\\
        J225549	& 343.95793	&  34.02002	& 0.51445(71)  &  56.0$\pm$4.1	 &  81.1$\pm$3.6   & 1.45(12) & 0.009(2)  &	0.028(4)\\
        J225630	& 344.12874	&  33.92002	& 0.41928(47)  &  60.8$\pm$3.0	 & 231.9$\pm$6.4   & 3.81(22) & 0.010(1)  &	0.542(45)\\
        J225840	& 344.66858	&  34.62947	& 0.60748(100) &  73.9$\pm$5.9	 & 116.3$\pm$7.1   & 1.57(16) &	0.025(6)  & 0.099(18)\\
        J225911	& 344.79647	&  36.35479	& 0.33796(41)  &  78.1$\pm$6.6	 & 100.4$\pm$12.8  & 1.29(20) & 0.017(4)  &	0.035(14)\\
        J230015	& 345.06424	&  32.5593	& 0.30691(34)  & 140.8$\pm$15.7	 & 157.5$\pm$13.4  & 1.12(16) & 0.089(30) & 0.124(32)\\
        J230252	& 345.71927	&  34.38347	& 0.60094(97)  &  56.8$\pm$5.3	 &  61.6$\pm$4.6   & 1.08(13) & 0.011(3)  & 0.015(3)\\
        J230446	& 346.19490	&  33.06033	& 0.51540(720) &  35.5$\pm$4.6	 &  42.0$\pm$2.9   & 1.18(17) & 0.002(1)  & 0.004(1)\\
        J234359	& 355.99732	&   0.45373	& 0.38337(60)  &  45.3$\pm$8.3	 & 105.7$\pm$12.5  & 2.33(51) & 0.004(2)  &	0.047(17)\\
        J234758	& 356.99417	&  39.92033	& 0.36709(35)  &  67.1$\pm$4.9	 &  75.2$\pm$6.1   & 1.12(12) & 0.011(3)  &	0.016(4)\\
        
        \hline
    \end{tabular}
    \caption{$K_1$ and $K_2$ denote the RV semi-amplitudes of the primary ($M_1$) and secondary ($M_2$), respectively; q = $M_2 / M_1$ is the mass ratio. The mass functions $f(M_1)$ and $f(M_2)$ are also derived. (This table is available in its entirety in machine-readable form.)
}
    \label{tab:rv_appendix}
\end{table}

\clearpage
\bibliography{ms.bib}{}

@misc{zhang2025,
  author       = {Zhang, Zhi-Xiang},
  title        = {zhang-zhixiang/stellarSpecModel: stellarSpecModel
                   release
                  },
  month        = mar,
  year         = 2025,
  publisher    = {Zenodo},
  version      = {v1.0.0},
  doi          = {10.5281/zenodo.15109587},
  url          = {https://doi.org/10.5281/zenodo.15109587},
  swhid        = {swh:1:dir:16c37119e5f7c2eb18a4c17d5fd49655658a31fc
                   ;origin=https://doi.org/10.5281/zenodo.15109586;vi
                   sit=swh:1:snp:b0422bac54352f0670435603beccaa55bcbb
                   3b1c;anchor=swh:1:rel:8ae122b78007860c8ef386f08a73
                   7e7a60927af6;path=zhang-zhixiang-
                   stellarSpecModel-9748279
                  },
}

@ARTICLE{Gustafsson2008,
       author = {{Gustafsson}, B. and {Edvardsson}, B. and {Eriksson}, K. and {J{\o}rgensen}, U.~G. and {Nordlund}, {\r{A}}. and {Plez}, B.},
        title = "{A grid of MARCS model atmospheres for late-type stars. I. Methods and general properties}",
      journal = {\aap},
         year = 2008,
        month = aug,
       volume = {486},
       number = {3},
        pages = {951-970},
          doi = {10.1051/0004-6361:200809724},
archivePrefix = {arXiv},
       eprint = {0805.0554},
 primaryClass = {astro-ph},
       adsurl = {https://ui.adsabs.harvard.edu/abs/2008A&A...486..951G}
}

@ARTICLE{elbadry2018,
       author = {{El-Badry}, Kareem and {Ting}, Yuan-Sen and {Rix}, Hans-Walter and {Quataert}, Eliot and {Weisz}, Daniel R. and {Cargile}, Phillip and {Conroy}, Charlie and {Hogg}, David W. and {Bergemann}, Maria and {Liu}, Chao},
        title = "{Discovery and characterization of 3000+ main-sequence binaries from APOGEE spectra}",
      journal = {\mnras},
         year = 2018,
        month = may,
       volume = {476},
       number = {1},
        pages = {528-553},
          doi = {10.1093/mnras/sty240},
archivePrefix = {arXiv},
       eprint = {1711.08793},
 primaryClass = {astro-ph.SR},
       adsurl = {https://ui.adsabs.harvard.edu/abs/2018MNRAS.476..528E}
}

@ARTICLE{Gonz2006,
       author = {{Gonz{\'a}lez}, J.~F. and {Levato}, H.},
        title = "{Separation of composite spectra: the spectroscopic detection of an eclipsing binary star}",
      journal = {\aap},
         year = 2006,
        month = mar,
       volume = {448},
       number = {1},
        pages = {283-292},
          doi = {10.1051/0004-6361:20053177},
       adsurl = {https://ui.adsabs.harvard.edu/abs/2006A&A...448..283G}
}

@ARTICLE{Zucker1994,
       author = {{Zucker}, S. and {Mazeh}, T.},
        title = "{Study of Spectroscopic Binaries with TODCOR. I. A New Two-dimensional Correlation Algorithm to Derive the Radial Velocities of the Two Components}",
      journal = {\apj},
         year = 1994,
        month = jan,
       volume = {420},
        pages = {806},
          doi = {10.1086/173605},
       adsurl = {https://ui.adsabs.harvard.edu/abs/1994ApJ...420..806Z}
}

@BOOK{1959K,
       author = {{Kopal}, Zdenek},
        title = "{Close binary systems}",
         year = 1959,
       adsurl = {https://ui.adsabs.harvard.edu/abs/1959cbs..book.....K}
}

@ARTICLE{1979ApJ...231..502L,
       author = {{Lucy}, L.~B. and {Wilson}, R.~E.},
        title = "{Observational tests of theories of contact binaries.}",
      journal = {\apj},
         year = 1979,
        month = jul,
       volume = {231},
        pages = {502-513},
          doi = {10.1086/157212},
       adsurl = {https://ui.adsabs.harvard.edu/abs/1979ApJ...231..502L}
}

@ARTICLE{1970VA.....12..217B,
       author = {{Binnendijk}, L.},
        title = "{The orbital elements of W Ursae Majoris systems}",
      journal = {Vistas in Astronomy},
         year = 1970,
        month = jan,
       volume = {12},
       number = {1},
        pages = {217-256},
          doi = {10.1016/0083-6656(70)90041-3},
       adsurl = {https://ui.adsabs.harvard.edu/abs/1970VA.....12..217B}
}

@ARTICLE{2017PhRvL.119p1101A,
       author = {{Abbott}, B.~P. and {Abbott}, R. and {Abbott}, T.~D. and {Acernese}, F. and {Ackley}, K. and {Adams}, C. and {Adams}, T. and {Addesso}, P. and {Adhikari}, R.~X. and {Adya}, V.~B. and {Affeldt}, C. and {Afrough}, M. and {Agarwal}, B. and {Agathos}, M. and {Agatsuma}, K. and {Aggarwal}, N. and {Aguiar}, O.~D. and {Aiello}, L. and {Ain}, A. and {Ajith}, P. and {Allen}, B. and {Allen}, G. and {Allocca}, A. and {Altin}, P.~A. and {Amato}, A. and {Ananyeva}, A. and {Anderson}, S.~B. and {Anderson}, W.~G. and {Angelova}, S.~V. and {Antier}, S. and {Appert}, S. and {Arai}, K. and {Araya}, M.~C. and {Areeda}, J.~S. and {Arnaud}, N. and {Arun}, K.~G. and {Ascenzi}, S. and {Ashton}, G. and {Ast}, M. and {Aston}, S.~M. and {Astone}, P. and {Atallah}, D.~V. and {Aufmuth}, P. and {Aulbert}, C. and {AultONeal}, K. and {Austin}, C. and {Avila-Alvarez}, A. and {Babak}, S. and {Bacon}, P. and {Bader}, M.~K.~M. and {Bae}, S. and {Bailes}, M. and {Baker}, P.~T. and {Baldaccini}, F. and {Ballardin}, G. and {Ballmer}, S.~W. and {Banagiri}, S. and {Barayoga}, J.~C. and {Barclay}, S.~E. and {Barish}, B.~C. and {Barker}, D. and {Barkett}, K. and {Barone}, F. and {Barr}, B. and {Barsotti}, L. and {Barsuglia}, M. and {Barta}, D. and {Barthelmy}, S.~D. and {Bartlett}, J. and {Bartos}, I. and {Bassiri}, R. and {Basti}, A. and {Batch}, J.~C. and {Bawaj}, M. and {Bayley}, J.~C. and {Bazzan}, M. and {B{\'e}csy}, B. and {Beer}, C. and {Bejger}, M. and {Belahcene}, I. and {Bell}, A.~S. and {Berger}, B.~K. and {Bergmann}, G. and {Bernuzzi}, S. and {Bero}, J.~J. and {Berry}, C.~P.~L. and {Bersanetti}, D. and {Bertolini}, A. and {Betzwieser}, J. and {Bhagwat}, S. and {Bhandare}, R. and {Bilenko}, I.~A. and {Billingsley}, G. and {Billman}, C.~R. and {Birch}, J. and {Birney}, R. and {Birnholtz}, O. and {Biscans}, S. and {Biscoveanu}, S. and {Bisht}, A. and {Bitossi}, M. and {Biwer}, C. and {Bizouard}, M.~A. and {Blackburn}, J.~K. and {Blackman}, J. and {Blair}, C.~D. and {Blair}, D.~G. and {Blair}, R.~M. and {Bloemen}, S. and {Bock}, O. and {Bode}, N. and {Boer}, M. and {Bogaert}, G. and {Bohe}, A. and {Bondu}, F. and {Bonilla}, E. and {Bonnand}, R. and {Boom}, B.~A. and {Bork}, R. and {Boschi}, V. and {Bose}, S. and {Bossie}, K. and {Bouffanais}, Y. and {Bozzi}, A. and {Bradaschia}, C. and {Brady}, P.~R. and {Branchesi}, M. and {Brau}, J.~E. and {Briant}, T. and {Brillet}, A. and {Brinkmann}, M. and {Brisson}, V. and {Brockill}, P. and {Broida}, J.~E. and {Brooks}, A.~F. and {Brown}, D.~A. and {Brown}, D.~D. and {Brunett}, S. and {Buchanan}, C.~C. and {Buikema}, A. and {Bulik}, T. and {Bulten}, H.~J. and {Buonanno}, A. and {Buskulic}, D. and {Buy}, C. and {Byer}, R.~L. and {Cabero}, M. and {Cadonati}, L. and {Cagnoli}, G. and {Cahillane}, C. and {Calder{\'o}n Bustillo}, J. and {Callister}, T.~A. and {Calloni}, E. and {Camp}, J.~B. and {Canepa}, M. and {Canizares}, P. and {Cannon}, K.~C. and {Cao}, H. and {Cao}, J. and {Capano}, C.~D. and {Capocasa}, E. and {Carbognani}, F. and {Caride}, S. and {Carney}, M.~F. and {Carullo}, G. and {Casanueva Diaz}, J. and {Casentini}, C. and {Caudill}, S. and {Cavagli{\`a}}, M. and {Cavalier}, F. and {Cavalieri}, R. and {Cella}, G. and {Cepeda}, C.~B. and {Cerd{\'a}-Dur{\'a}n}, P. and {Cerretani}, G. and {Cesarini}, E. and {Chamberlin}, S.~J. and {Chan}, M. and {Chao}, S. and {Charlton}, P. and {Chase}, E. and {Chassande-Mottin}, E. and {Chatterjee}, D. and {Chatziioannou}, K. and {Cheeseboro}, B.~D. and {Chen}, H.~Y. and {Chen}, X. and {Chen}, Y. and {Cheng}, H. -P. and {Chia}, H. and {Chincarini}, A. and {Chiummo}, A. and {Chmiel}, T. and {Cho}, H.~S. and {Cho}, M. and {Chow}, J.~H. and {Christensen}, N. and {Chu}, Q. and {Chua}, A.~J.~K. and {Chua}, S. and {Chung}, A.~K.~W. and {Chung}, S. and {Ciani}, G. and {Ciolfi}, R. and {Cirelli}, C.~E. and {Cirone}, A. and {Clara}, F. and {Clark}, J.~A. and {Clearwater}, P. and {Cleva}, F. and {Cocchieri}, C. and {Coccia}, E. and {Cohadon}, P. -F. and {Cohen}, D. and {Colla}, A. and {Collette}, C.~G. and {Cominsky}, L.~R. and {Constancio}, M. and {Conti}, L. and {Cooper}, S.~J. and {Corban}, P. and {Corbitt}, T.~R. and {Cordero-Carri{\'o}n}, I. and {Corley}, K.~R. and {Cornish}, N. and {Corsi}, A. and {Cortese}, S. and {Costa}, C.~A. and {Coughlin}, M.~W. and {Coughlin}, S.~B. and {Coulon}, J. -P. and {Countryman}, S.~T. and {Couvares}, P. and {Covas}, P.~B. and {Cowan}, E.~E. and {Coward}, D.~M. and {Cowart}, M.~J. and {Coyne}, D.~C. and {Coyne}, R. and {Creighton}, J.~D.~E. and {Creighton}, T.~D. and {Cripe}, J. and {Crowder}, S.~G. and {Cullen}, T.~J. and {Cumming}, A. and {Cunningham}, L. and {Cuoco}, E. and {Dal Canton}, T. and {D{\'a}lya}, G. and {Danilishin}, S.~L. and {D'Antonio}, S. and {Danzmann}, K. and {Dasgupta}, A. and {Da Silva Costa}, C.~F. and {Dattilo}, V. and {Dave}, I. and {Davier}, M. and {Davis}, D. and {Daw}, E.~J. and {Day}, B. and {De}, S. and {DeBra}, D. and {Degallaix}, J. and {De Laurentis}, M. and {Del{\'e}glise}, S. and {Del Pozzo}, W. and {Demos}, N. and {Denker}, T. and {Dent}, T. and {De Pietri}, R. and {Dergachev}, V. and {De Rosa}, R. and {DeRosa}, R.~T. and {De Rossi}, C. and {DeSalvo}, R. and {de Varona}, O. and {Devenson}, J. and {Dhurandhar}, S. and {D{\'\i}az}, M.~C. and {Dietrich}, T. and {Di Fiore}, L. and {Di Giovanni}, M. and {Di Girolamo}, T. and {Di Lieto}, A. and {Di Pace}, S. and {Di Palma}, I. and {Di Renzo}, F. and {Doctor}, Z. and {Dolique}, V. and {Donovan}, F. and {Dooley}, K.~L. and {Doravari}, S. and {Dorrington}, I. and {Douglas}, R. and {Dovale {\'A}lvarez}, M. and {Downes}, T.~P. and {Drago}, M. and {Dreissigacker}, C. and {Driggers}, J.~C. and {Du}, Z. and {Ducrot}, M. and {Dudi}, R. and {Dupej}, P. and {Dwyer}, S.~E. and {Edo}, T.~B. and {Edwards}, M.~C. and {Effler}, A. and {Eggenstein}, H. -B. and {Ehrens}, P. and {Eichholz}, J. and {Eikenberry}, S.~S. and {Eisenstein}, R.~A. and {Essick}, R.~C. and {Estevez}, D. and {Etienne}, Z.~B. and {Etzel}, T. and {Evans}, M. and {Evans}, T.~M. and {Factourovich}, M. and {Fafone}, V. and {Fair}, H. and {Fairhurst}, S. and {Fan}, X. and {Farinon}, S. and {Farr}, B. and {Farr}, W.~M. and {Fauchon-Jones}, E.~J. and {Favata}, M. and {Fays}, M. and {Fee}, C. and {Fehrmann}, H. and {Feicht}, J. and {Fejer}, M.~M. and {Fernandez-Galiana}, A. and {Ferrante}, I. and {Ferreira}, E.~C. and {Ferrini}, F. and {Fidecaro}, F. and {Finstad}, D. and {Fiori}, I. and {Fiorucci}, D. and {Fishbach}, M. and {Fisher}, R.~P. and {Fitz-Axen}, M. and {Flaminio}, R. and {Fletcher}, M. and {Fong}, H. and {Font}, J.~A. and {Forsyth}, P.~W.~F. and {Forsyth}, S.~S. and {Fournier}, J. -D. and {Frasca}, S. and {Frasconi}, F. and {Frei}, Z. and {Freise}, A. and {Frey}, R. and {Frey}, V. and {Fries}, E.~M. and {Fritschel}, P. and {Frolov}, V.~V. and {Fulda}, P. and {Fyffe}, M. and {Gabbard}, H. and {Gadre}, B.~U. and {Gaebel}, S.~M. and {Gair}, J.~R. and {Gammaitoni}, L. and {Ganija}, M.~R. and {Gaonkar}, S.~G. and {Garcia-Quiros}, C. and {Garufi}, F. and {Gateley}, B. and {Gaudio}, S. and {Gaur}, G. and {Gayathri}, V. and {Gehrels}, N. and {Gemme}, G. and {Genin}, E. and {Gennai}, A. and {George}, D. and {George}, J. and {Gergely}, L. and {Germain}, V. and {Ghonge}, S. and {Ghosh}, Abhirup and {Ghosh}, Archisman and {Ghosh}, S. and {Giaime}, J.~A. and {Giardina}, K.~D. and {Giazotto}, A. and {Gill}, K. and {Glover}, L. and {Goetz}, E. and {Goetz}, R. and {Gomes}, S. and {Goncharov}, B. and {Gonz{\'a}lez}, G. and {Gonzalez Castro}, J.~M. and {Gopakumar}, A. and {Gorodetsky}, M.~L. and {Gossan}, S.~E. and {Gosselin}, M. and {Gouaty}, R. and {Grado}, A. and {Graef}, C. and {Granata}, M. and {Grant}, A. and {Gras}, S. and {Gray}, C. and {Greco}, G. and {Green}, A.~C. and {Gretarsson}, E.~M. and {Groot}, P. and {Grote}, H. and {Grunewald}, S. and {Gruning}, P. and {Guidi}, G.~M. and {Guo}, X. and {Gupta}, A. and {Gupta}, M.~K. and {Gushwa}, K.~E. and {Gustafson}, E.~K. and {Gustafson}, R. and {Halim}, O. and {Hall}, B.~R. and {Hall}, E.~D. and {Hamilton}, E.~Z. and {Hammond}, G. and {Haney}, M. and {Hanke}, M.~M. and {Hanks}, J. and {Hanna}, C. and {Hannam}, M.~D. and {Hannuksela}, O.~A. and {Hanson}, J. and {Hardwick}, T. and {Harms}, J. and {Harry}, G.~M. and {Harry}, I.~W. and {Hart}, M.~J. and {Haster}, C. -J. and {Haughian}, K. and {Healy}, J. and {Heidmann}, A. and {Heintze}, M.~C. and {Heitmann}, H. and {Hello}, P. and {Hemming}, G. and {Hendry}, M. and {Heng}, I.~S. and {Hennig}, J. and {Heptonstall}, A.~W. and {Heurs}, M. and {Hild}, S. and {Hinderer}, T. and {Ho}, W.~C.~G. and {Hoak}, D. and {Hofman}, D. and {Holt}, K. and {Holz}, D.~E. and {Hopkins}, P. and {Horst}, C. and {Hough}, J. and {Houston}, E.~A. and {Howell}, E.~J. and {Hreibi}, A. and {Hu}, Y.~M. and {Huerta}, E.~A. and {Huet}, D. and {Hughey}, B. and {Husa}, S. and {Huttner}, S.~H. and {Huynh-Dinh}, T. and {Indik}, N. and {Inta}, R. and {Intini}, G. and {Isa}, H.~N. and {Isac}, J. -M. and {Isi}, M. and {Iyer}, B.~R. and {Izumi}, K. and {Jacqmin}, T. and {Jani}, K. and {Jaranowski}, P. and {Jawahar}, S. and {Jim{\'e}nez-Forteza}, F. and {Johnson}, W.~W. and {Johnson-McDaniel}, N.~K. and {Jones}, D.~I. and {Jones}, R. and {Jonker}, R.~J.~G. and {Ju}, L. and {Junker}, J. and {Kalaghatgi}, C.~V. and {Kalogera}, V. and {Kamai}, B. and {Kandhasamy}, S. and {Kang}, G. and {Kanner}, J.~B. and {Kapadia}, S.~J. and {Karki}, S. and {Karvinen}, K.~S. and {Kasprzack}, M. and {Kastaun}, W. and {Katolik}, M. and {Katsavounidis}, E. and {Katzman}, W. and {Kaufer}, S. and {Kawabe}, K. and {K{\'e}f{\'e}lian}, F. and {Keitel}, D. and {Kemball}, A.~J. and {Kennedy}, R. and {Kent}, C. and {Key}, J.~S. and {Khalili}, F.~Y. and {Khan}, I. and {Khan}, S. and {Khan}, Z. and {Khazanov}, E.~A. and {Kijbunchoo}, N. and {Kim}, Chunglee and {Kim}, J.~C. and {Kim}, K. and {Kim}, W. and {Kim}, W.~S. and {Kim}, Y. -M. and {Kimbrell}, S.~J. and {King}, E.~J. and {King}, P.~J. and {Kinley-Hanlon}, M. and {Kirchhoff}, R. and {Kissel}, J.~S. and {Kleybolte}, L. and {Klimenko}, S. and {Knowles}, T.~D. and {Koch}, P. and {Koehlenbeck}, S.~M. and {Koley}, S. and {Kondrashov}, V. and {Kontos}, A. and {Korobko}, M. and {Korth}, W.~Z. and {Kowalska}, I. and {Kozak}, D.~B. and {Kr{\"a}mer}, C. and {Kringel}, V. and {Krishnan}, B. and {Kr{\'o}lak}, A. and {Kuehn}, G. and {Kumar}, P. and {Kumar}, R. and {Kumar}, S. and {Kuo}, L. and {Kutynia}, A. and {Kwang}, S. and {Lackey}, B.~D. and {Lai}, K.~H. and {Landry}, M. and {Lang}, R.~N. and {Lange}, J. and {Lantz}, B. and {Lanza}, R.~K. and {Larson}, S.~L. and {Lartaux-Vollard}, A. and {Lasky}, P.~D. and {Laxen}, M. and {Lazzarini}, A. and {Lazzaro}, C. and {Leaci}, P. and {Leavey}, S. and {Lee}, C.~H. and {Lee}, H.~K. and {Lee}, H.~M. and {Lee}, H.~W. and {Lee}, K. and {Lehmann}, J. and {Lenon}, A. and {Leon}, E. and {Leonardi}, M. and {Leroy}, N. and {Letendre}, N. and {Levin}, Y. and {Li}, T.~G.~F. and {Linker}, S.~D. and {Littenberg}, T.~B. and {Liu}, J. and {Liu}, X. and {Lo}, R.~K.~L. and {Lockerbie}, N.~A. and {London}, L.~T. and {Lord}, J.~E. and {Lorenzini}, M. and {Loriette}, V. and {Lormand}, M. and {Losurdo}, G. and {Lough}, J.~D. and {Lousto}, C.~O. and {Lovelace}, G. and {L{\"u}ck}, H. and {Lumaca}, D. and {Lundgren}, A.~P. and {Lynch}, R. and {Ma}, Y. and {Macas}, R. and {Macfoy}, S. and {Machenschalk}, B. and {MacInnis}, M. and {Macleod}, D.~M. and {Maga{\~n}a Hernandez}, I. and {Maga{\~n}a-Sandoval}, F. and {Maga{\~n}a Zertuche}, L. and {Magee}, R.~M. and {Majorana}, E. and {Maksimovic}, I. and {Man}, N. and {Mandic}, V. and {Mangano}, V. and {Mansell}, G.~L. and {Manske}, M. and {Mantovani}, M. and {Marchesoni}, F. and {Marion}, F. and {M{\'a}rka}, S. and {M{\'a}rka}, Z. and {Markakis}, C. and {Markosyan}, A.~S. and {Markowitz}, A. and {Maros}, E. and {Marquina}, A. and {Marsh}, P. and {Martelli}, F. and {Martellini}, L. and {Martin}, I.~W. and {Martin}, R.~M. and {Martynov}, D.~V. and {Marx}, J.~N. and {Mason}, K. and {Massera}, E. and {Masserot}, A. and {Massinger}, T.~J. and {Masso-Reid}, M. and {Mastrogiovanni}, S. and {Matas}, A. and {Matichard}, F. and {Matone}, L. and {Mavalvala}, N. and {Mazumder}, N. and {McCarthy}, R. and {McClelland}, D.~E. and {McCormick}, S. and {McCuller}, L. and {McGuire}, S.~C. and {McIntyre}, G. and {McIver}, J. and {McManus}, D.~J. and {McNeill}, L. and {McRae}, T. and {McWilliams}, S.~T. and {Meacher}, D. and {Meadors}, G.~D. and {Mehmet}, M. and {Meidam}, J. and {Mejuto-Villa}, E. and {Melatos}, A. and {Mendell}, G. and {Mercer}, R.~A. and {Merilh}, E.~L. and {Merzougui}, M. and {Meshkov}, S. and {Messenger}, C. and {Messick}, C. and {Metzdorff}, R. and {Meyers}, P.~M. and {Miao}, H. and {Michel}, C. and {Middleton}, H. and {Mikhailov}, E.~E. and {Milano}, L. and {Miller}, A.~L. and {Miller}, B.~B. and {Miller}, J. and {Millhouse}, M. and {Milovich-Goff}, M.~C. and {Minazzoli}, O. and {Minenkov}, Y. and {Ming}, J. and {Mishra}, C. and {Mitra}, S. and {Mitrofanov}, V.~P. and {Mitselmakher}, G. and {Mittleman}, R. and {Moffa}, D. and {Moggi}, A. and {Mogushi}, K. and {Mohan}, M. and {Mohapatra}, S.~R.~P. and {Molina}, I. and {Montani}, M. and {Moore}, C.~J. and {Moraru}, D. and {Moreno}, G. and {Morisaki}, S. and {Morriss}, S.~R. and {Mours}, B. and {Mow-Lowry}, C.~M. and {Mueller}, G. and {Muir}, A.~W. and {Mukherjee}, Arunava and {Mukherjee}, D. and {Mukherjee}, S. and {Mukund}, N. and {Mullavey}, A. and {Munch}, J. and {Mu{\~n}iz}, E.~A. and {Muratore}, M. and {Murray}, P.~G. and {Nagar}, A. and {Napier}, K. and {Nardecchia}, I. and {Naticchioni}, L. and {Nayak}, R.~K. and {Neilson}, J. and {Nelemans}, G. and {Nelson}, T.~J.~N. and {Nery}, M. and {Neunzert}, A. and {Nevin}, L. and {Newport}, J.~M. and {Newton}, G. and {Ng}, K.~K.~Y. and {Nguyen}, P. and {Nguyen}, T.~T. and {Nichols}, D. and {Nielsen}, A.~B. and {Nissanke}, S. and {Nitz}, A. and {Noack}, A. and {Nocera}, F. and {Nolting}, D. and {North}, C. and {Nuttall}, L.~K. and {Oberling}, J. and {O'Dea}, G.~D. and {Ogin}, G.~H. and {Oh}, J.~J. and {Oh}, S.~H. and {Ohme}, F. and {Okada}, M.~A. and {Oliver}, M. and {Oppermann}, P. and {Oram}, Richard J. and {O'Reilly}, B. and {Ormiston}, R. and {Ortega}, L.~F. and {O'Shaughnessy}, R. and {Ossokine}, S. and {Ottaway}, D.~J. and {Overmier}, H. and {Owen}, B.~J. and {Pace}, A.~E. and {Page}, J. and {Page}, M.~A. and {Pai}, A. and {Pai}, S.~A. and {Palamos}, J.~R. and {Palashov}, O. and {Palomba}, C. and {Pal-Singh}, A. and {Pan}, Howard and {Pan}, Huang-Wei and {Pang}, B. and {Pang}, P.~T.~H. and {Pankow}, C. and {Pannarale}, F. and {Pant}, B.~C. and {Paoletti}, F. and {Paoli}, A. and {Papa}, M.~A. and {Parida}, A. and {Parker}, W. and {Pascucci}, D. and {Pasqualetti}, A. and {Passaquieti}, R. and {Passuello}, D. and {Patil}, M. and {Patricelli}, B. and {Pearlstone}, B.~L. and {Pedraza}, M. and {Pedurand}, R. and {Pekowsky}, L. and {Pele}, A. and {Penn}, S. and {Perez}, C.~J. and {Perreca}, A. and {Perri}, L.~M. and {Pfeiffer}, H.~P. and {Phelps}, M. and {Piccinni}, O.~J. and {Pichot}, M. and {Piergiovanni}, F. and {Pierro}, V. and {Pillant}, G. and {Pinard}, L. and {Pinto}, I.~M. and {Pirello}, M. and {Pitkin}, M. and {Poe}, M. and {Poggiani}, R. and {Popolizio}, P. and {Porter}, E.~K. and {Post}, A. and {Powell}, J. and {Prasad}, J. and {Pratt}, J.~W.~W. and {Pratten}, G. and {Predoi}, V. and {Prestegard}, T. and {Prijatelj}, M. and {Principe}, M. and {Privitera}, S. and {Prix}, R. and {Prodi}, G.~A. and {Prokhorov}, L.~G. and {Puncken}, O. and {Punturo}, M. and {Puppo}, P. and {P{\"u}rrer}, M. and {Qi}, H. and {Quetschke}, V. and {Quintero}, E.~A. and {Quitzow-James}, R. and {Raab}, F.~J. and {Rabeling}, D.~S. and {Radkins}, H. and {Raffai}, P. and {Raja}, S. and {Rajan}, C. and {Rajbhandari}, B. and {Rakhmanov}, M. and {Ramirez}, K.~E. and {Ramos-Buades}, A. and {Rapagnani}, P. and {Raymond}, V. and {Razzano}, M. and {Read}, J. and {Regimbau}, T. and {Rei}, L. and {Reid}, S. and {Reitze}, D.~H. and {Ren}, W. and {Reyes}, S.~D. and {Ricci}, F. and {Ricker}, P.~M. and {Rieger}, S. and {Riles}, K. and {Rizzo}, M. and {Robertson}, N.~A. and {Robie}, R. and {Robinet}, F. and {Rocchi}, A. and {Rolland}, L. and {Rollins}, J.~G. and {Roma}, V.~J. and {Romano}, J.~D. and {Romano}, R. and {Romel}, C.~L. and {Romie}, J.~H. and {Rosi{\'n}ska}, D. and {Ross}, M.~P. and {Rowan}, S. and {R{\"u}diger}, A. and {Ruggi}, P. and {Rutins}, G. and {Ryan}, K. and {Sachdev}, S. and {Sadecki}, T. and {Sadeghian}, L. and {Sakellariadou}, M. and {Salconi}, L. and {Saleem}, M. and {Salemi}, F. and {Samajdar}, A. and {Sammut}, L. and {Sampson}, L.~M. and {Sanchez}, E.~J. and {Sanchez}, L.~E. and {Sanchis-Gual}, N. and {Sandberg}, V. and {Sanders}, J.~R. and {Sassolas}, B. and {Sathyaprakash}, B.~S. and {Saulson}, P.~R. and {Sauter}, O. and {Savage}, R.~L. and {Sawadsky}, A. and {Schale}, P. and {Scheel}, M. and {Scheuer}, J. and {Schmidt}, J. and {Schmidt}, P. and {Schnabel}, R. and {Schofield}, R.~M.~S. and {Sch{\"o}nbeck}, A. and {Schreiber}, E. and {Schuette}, D. and {Schulte}, B.~W. and {Schutz}, B.~F. and {Schwalbe}, S.~G. and {Scott}, J. and {Scott}, S.~M. and {Seidel}, E. and {Sellers}, D. and {Sengupta}, A.~S. and {Sentenac}, D. and {Sequino}, V. and {Sergeev}, A. and {Shaddock}, D.~A. and {Shaffer}, T.~J. and {Shah}, A.~A. and {Shahriar}, M.~S. and {Shaner}, M.~B. and {Shao}, L. and {Shapiro}, B. and {Shawhan}, P. and {Sheperd}, A. and {Shoemaker}, D.~H. and {Shoemaker}, D.~M. and {Siellez}, K. and {Siemens}, X. and {Sieniawska}, M. and {Sigg}, D. and {Silva}, A.~D. and {Singer}, L.~P. and {Singh}, A. and {Singhal}, A. and {Sintes}, A.~M. and {Slagmolen}, B.~J.~J. and {Smith}, B. and {Smith}, J.~R. and {Smith}, R.~J.~E. and {Somala}, S. and {Son}, E.~J. and {Sonnenberg}, J.~A. and {Sorazu}, B. and {Sorrentino}, F. and {Souradeep}, T. and {Spencer}, A.~P. and {Srivastava}, A.~K. and {Staats}, K. and {Staley}, A. and {Steinke}, M. and {Steinlechner}, J. and {Steinlechner}, S. and {Steinmeyer}, D. and {Stevenson}, S.~P. and {Stone}, R. and {Stops}, D.~J. and {Strain}, K.~A. and {Stratta}, G. and {Strigin}, S.~E. and {Strunk}, A. and {Sturani}, R. and {Stuver}, A.~L. and {Summerscales}, T.~Z. and {Sun}, L. and {Sunil}, S. and {Suresh}, J. and {Sutton}, P.~J. and {Swinkels}, B.~L. and {Szczepa{\'n}czyk}, M.~J. and {Tacca}, M. and {Tait}, S.~C. and {Talbot}, C. and {Talukder}, D. and {Tanner}, D.~B. and {T{\'a}pai}, M. and {Taracchini}, A. and {Tasson}, J.~D. and {Taylor}, J.~A. and {Taylor}, R. and {Tewari}, S.~V. and {Theeg}, T. and {Thies}, F. and {Thomas}, E.~G. and {Thomas}, M. and {Thomas}, P. and {Thorne}, K.~A. and {Thorne}, K.~S. and {Thrane}, E. and {Tiwari}, S. and {Tiwari}, V. and {Tokmakov}, K.~V. and {Toland}, K. and {Tonelli}, M. and {Tornasi}, Z. and {Torres-Forn{\'e}}, A. and {Torrie}, C.~I. and {T{\"o}yr{\"a}}, D. and {Travasso}, F. and {Traylor}, G. and {Trinastic}, J. and {Tringali}, M.~C. and {Trozzo}, L. and {Tsang}, K.~W. and {Tse}, M. and {Tso}, R. and {Tsukada}, L. and {Tsuna}, D. and {Tuyenbayev}, D. and {Ueno}, K. and {Ugolini}, D. and {Unnikrishnan}, C.~S. and {Urban}, A.~L. and {Usman}, S.~A. and {Vahlbruch}, H. and {Vajente}, G. and {Valdes}, G. and {Vallisneri}, M. and {van Bakel}, N. and {van Beuzekom}, M. and {van den Brand}, J.~F.~J. and {Van Den Broeck}, C. and {Vander-Hyde}, D.~C. and {van der Schaaf}, L. and {van Heijningen}, J.~V. and {van Veggel}, A.~A. and {Vardaro}, M. and {Varma}, V. and {Vass}, S. and {Vas{\'u}th}, M. and {Vecchio}, A. and {Vedovato}, G. and {Veitch}, J. and {Veitch}, P.~J. and {Venkateswara}, K. and {Venugopalan}, G. and {Verkindt}, D. and {Vetrano}, F. and {Vicer{\'e}}, A. and {Viets}, A.~D. and {Vinciguerra}, S. and {Vine}, D.~J. and {Vinet}, J. -Y. and {Vitale}, S. and {Vo}, T. and {Vocca}, H. and {Vorvick}, C. and {Vyatchanin}, S.~P. and {Wade}, A.~R. and {Wade}, L.~E. and {Wade}, M. and {Walet}, R. and {Walker}, M. and {Wallace}, L. and {Walsh}, S. and {Wang}, G. and {Wang}, H. and {Wang}, J.~Z. and {Wang}, W.~H. and {Wang}, Y.~F. and {Ward}, R.~L. and {Warner}, J. and {Was}, M. and {Watchi}, J. and {Weaver}, B. and {Wei}, L. -W. and {Weinert}, M. and {Weinstein}, A.~J. and {Weiss}, R. and {Wen}, L. and {Wessel}, E.~K. and {We{\ss}els}, P. and {Westerweck}, J. and {Westphal}, T. and {Wette}, K. and {Whelan}, J.~T. and {Whitcomb}, S.~E. and {Whiting}, B.~F. and {Whittle}, C. and {Wilken}, D. and {Williams}, D. and {Williams}, R.~D. and {Williamson}, A.~R. and {Willis}, J.~L. and {Willke}, B. and {Wimmer}, M.~H. and {Winkler}, W. and {Wipf}, C.~C. and {Wittel}, H. and {Woan}, G. and {Woehler}, J. and {Wofford}, J. and {Wong}, K.~W.~K. and {Worden}, J. and {Wright}, J.~L. and {Wu}, D.~S. and {Wysocki}, D.~M. and {Xiao}, S. and {Yamamoto}, H. and {Yancey}, C.~C. and {Yang}, L. and {Yap}, M.~J. and {Yazback}, M. and {Yu}, Hang and {Yu}, Haocun and {Yvert}, M. and {Zadro{\.Z}ny}, A. and {Zanolin}, M. and {Zelenova}, T. and {Zendri}, J. -P. and {Zevin}, M. and {Zhang}, L. and {Zhang}, M. and {Zhang}, T. and {Zhang}, Y. -H. and {Zhao}, C. and {Zhou}, M. and {Zhou}, Z. and {Zhu}, S.~J. and {Zhu}, X.~J. and {Zimmerman}, A.~B. and {Zucker}, M.~E. and {Zweizig}, J. and {LIGO Scientific Collaboration} and {Virgo Collaboration}},
        title = "{GW170817: Observation of Gravitational Waves from a Binary Neutron Star Inspiral}",
      journal = {\prl},
         year = 2017,
        month = oct,
       volume = {119},
       number = {16},
          eid = {161101},
        pages = {161101},
          doi = {10.1103/PhysRevLett.119.161101},
archivePrefix = {arXiv},
       eprint = {1710.05832},
 primaryClass = {gr-qc},
       adsurl = {https://ui.adsabs.harvard.edu/abs/2017PhRvL.119p1101A}
}

@ARTICLE{2007A&A...467..229P,
       author = {{Panov}, K. and {Dimitrov}, D.},
        title = "{Long-term photometric study of <ASTROBJ>FK Comae</ASTROBJ> Berenices and <ASTROBJ>HD 199178</ASTROBJ>}",
      journal = {\aap},
         year = 2007,
        month = may,
       volume = {467},
       number = {1},
        pages = {229-235},
          doi = {10.1051/0004-6361:20065596},
       adsurl = {https://ui.adsabs.harvard.edu/abs/2007A&A...467..229P}
}

@ARTICLE{1987ApJ...314..585K,
       author = {{Kaluzny}, Janusz and {Shara}, Michael M.},
        title = "{The Discovery of Six New Short-Period Variables in the Old Open Cluster NGC 188}",
      journal = {\apj},
         year = 1987,
        month = mar,
       volume = {314},
        pages = {585},
          doi = {10.1086/165087},
       adsurl = {https://ui.adsabs.harvard.edu/abs/1987ApJ...314..585K}
}

@ARTICLE{2023A&ARv..31....1A,
       author = {{Arcones}, Almudena and {Thielemann}, Friedrich-Karl},
        title = "{Origin of the elements}",
      journal = {\aapr},
         year = 2023,
        month = dec,
       volume = {31},
       number = {1},
          eid = {1},
        pages = {1},
          doi = {10.1007/s00159-022-00146-x},
       adsurl = {https://ui.adsabs.harvard.edu/abs/2023A&ARv..31....1A}
}

@ARTICLE{2013Natur.495...76P,
       author = {{Pietrzy{\'n}ski}, G. and {Graczyk}, D. and {Gieren}, W. and {Thompson}, I.~B. and {Pilecki}, B. and {Udalski}, A. and {Soszy{\'n}ski}, I. and {Koz{\l}owski}, S. and {Konorski}, P. and {Suchomska}, K. and {Bono}, G. and {Moroni}, P.~G. Prada and {Villanova}, S. and {Nardetto}, N. and {Bresolin}, F. and {Kudritzki}, R.~P. and {Storm}, J. and {Gallenne}, A. and {Smolec}, R. and {Minniti}, D. and {Kubiak}, M. and {Szyma{\'n}ski}, M.~K. and {Poleski}, R. and {Wyrzykowski}, {\L}. and {Ulaczyk}, K. and {Pietrukowicz}, P. and {G{\'o}rski}, M. and {Karczmarek}, P.},
        title = "{An eclipsing-binary distance to the Large Magellanic Cloud accurate to two per cent}",
      journal = {\nat},
         year = 2013,
        month = mar,
       volume = {495},
       number = {7439},
        pages = {76-79},
          doi = {10.1038/nature11878},
archivePrefix = {arXiv},
       eprint = {1303.2063},
 primaryClass = {astro-ph.GA},
       adsurl = {https://ui.adsabs.harvard.edu/abs/2013Natur.495...76P}
}

@ARTICLE{2014MNRAS.445L..74M,
       author = {{McKernan}, B. and {Ford}, K.~E.~S. and {Kocsis}, B. and {Haiman}, Z.},
        title = "{Stars as resonant absorbers of gravitational waves.}",
      journal = {\mnras},
         year = 2014,
        month = nov,
       volume = {445},
        pages = {L74-L78},
          doi = {10.1093/mnrasl/slu136},
archivePrefix = {arXiv},
       eprint = {1405.1414},
 primaryClass = {astro-ph.HE},
       adsurl = {https://ui.adsabs.harvard.edu/abs/2014MNRAS.445L..74M}
}

@ARTICLE{2014ApJ...780...59G,
       author = {{Graczyk}, Dariusz and {Pietrzy{\'n}ski}, Grzegorz and {Thompson}, Ian B. and {Gieren}, Wolfgang and {Pilecki}, Bogumi{\l} and {Konorski}, Piotr and {Udalski}, Andrzej and {Soszy{\'n}ski}, Igor and {Villanova}, Sandro and {G{\'o}rski}, Marek and {Suchomska}, Ksenia and {Karczmarek}, Paulina and {Kudritzki}, Rolf-Peter and {Bresolin}, Fabio and {Gallenne}, Alexandre},
        title = "{The Araucaria Project. The Distance to the Small Magellanic Cloud from Late-type Eclipsing Binaries}",
      journal = {\apj},
         year = 2014,
        month = jan,
       volume = {780},
       number = {1},
          eid = {59},
        pages = {59},
          doi = {10.1088/0004-637X/780/1/59},
archivePrefix = {arXiv},
       eprint = {1311.2340},
 primaryClass = {astro-ph.CO},
       adsurl = {https://ui.adsabs.harvard.edu/abs/2014ApJ...780...59G}
}

@ARTICLE{2016PhRvL.116f1102A,
       author = {{Abbott}, B.~P. and {Abbott}, R. and {Abbott}, T.~D. and {Abernathy}, M.~R. and {Acernese}, F. and {Ackley}, K. and {Adams}, C. and {Adams}, T. and {Addesso}, P. and {Adhikari}, R.~X. and {Adya}, V.~B. and {Affeldt}, C. and {Agathos}, M. and {Agatsuma}, K. and {Aggarwal}, N. and {Aguiar}, O.~D. and {Aiello}, L. and {Ain}, A. and {Ajith}, P. and {Allen}, B. and {Allocca}, A. and {Altin}, P.~A. and {Anderson}, S.~B. and {Anderson}, W.~G. and {Arai}, K. and {Arain}, M.~A. and {Araya}, M.~C. and {Arceneaux}, C.~C. and {Areeda}, J.~S. and {Arnaud}, N. and {Arun}, K.~G. and {Ascenzi}, S. and {Ashton}, G. and {Ast}, M. and {Aston}, S.~M. and {Astone}, P. and {Aufmuth}, P. and {Aulbert}, C. and {Babak}, S. and {Bacon}, P. and {Bader}, M.~K.~M. and {Baker}, P.~T. and {Baldaccini}, F. and {Ballardin}, G. and {Ballmer}, S.~W. and {Barayoga}, J.~C. and {Barclay}, S.~E. and {Barish}, B.~C. and {Barker}, D. and {Barone}, F. and {Barr}, B. and {Barsotti}, L. and {Barsuglia}, M. and {Barta}, D. and {Bartlett}, J. and {Barton}, M.~A. and {Bartos}, I. and {Bassiri}, R. and {Basti}, A. and {Batch}, J.~C. and {Baune}, C. and {Bavigadda}, V. and {Bazzan}, M. and {Behnke}, B. and {Bejger}, M. and {Belczynski}, C. and {Bell}, A.~S. and {Bell}, C.~J. and {Berger}, B.~K. and {Bergman}, J. and {Bergmann}, G. and {Berry}, C.~P.~L. and {Bersanetti}, D. and {Bertolini}, A. and {Betzwieser}, J. and {Bhagwat}, S. and {Bhandare}, R. and {Bilenko}, I.~A. and {Billingsley}, G. and {Birch}, J. and {Birney}, I.~A. and {Birnholtz}, O. and {Biscans}, S. and {Bisht}, A. and {Bitossi}, M. and {Biwer}, C. and {Bizouard}, M.~A. and {Blackburn}, J.~K. and {Blair}, C.~D. and {Blair}, D.~G. and {Blair}, R.~M. and {Bloemen}, S. and {Bock}, O. and {Bodiya}, T.~P. and {Boer}, M. and {Bogaert}, G. and {Bogan}, C. and {Bohe}, A. and {Bojtos}, P. and {Bond}, C. and {Bondu}, F. and {Bonnand}, R. and {Boom}, B.~A. and {Bork}, R. and {Boschi}, V. and {Bose}, S. and {Bouffanais}, Y. and {Bozzi}, A. and {Bradaschia}, C. and {Brady}, P.~R. and {Braginsky}, V.~B. and {Branchesi}, M. and {Brau}, J.~E. and {Briant}, T. and {Brillet}, A. and {Brinkmann}, M. and {Brisson}, V. and {Brockill}, P. and {Brooks}, A.~F. and {Brown}, D.~A. and {Brown}, D.~D. and {Brown}, N.~M. and {Buchanan}, C.~C. and {Buikema}, A. and {Bulik}, T. and {Bulten}, H.~J. and {Buonanno}, A. and {Buskulic}, D. and {Buy}, C. and {Byer}, R.~L. and {Cabero}, M. and {Cadonati}, L. and {Cagnoli}, G. and {Cahillane}, C. and {Bustillo}, J. Calder{\'o}n and {Callister}, T. and {Calloni}, E. and {Camp}, J.~B. and {Cannon}, K.~C. and {Cao}, J. and {Capano}, C.~D. and {Capocasa}, E. and {Carbognani}, F. and {Caride}, S. and {Diaz}, J. Casanueva and {Casentini}, C. and {Caudill}, S. and {Cavagli{\`a}}, M. and {Cavalier}, F. and {Cavalieri}, R. and {Cella}, G. and {Cepeda}, C.~B. and {Baiardi}, L. Cerboni and {Cerretani}, G. and {Cesarini}, E. and {Chakraborty}, R. and {Chalermsongsak}, T. and {Chamberlin}, S.~J. and {Chan}, M. and {Chao}, S. and {Charlton}, P. and {Chassande-Mottin}, E. and {Chen}, H.~Y. and {Chen}, Y. and {Cheng}, C. and {Chincarini}, A. and {Chiummo}, A. and {Cho}, H.~S. and {Cho}, M. and {Chow}, J.~H. and {Christensen}, N. and {Chu}, Q. and {Chua}, S. and {Chung}, S. and {Ciani}, G. and {Clara}, F. and {Clark}, J.~A. and {Cleva}, F. and {Coccia}, E. and {Cohadon}, P. -F. and {Colla}, A. and {Collette}, C.~G. and {Cominsky}, L. and {Constancio}, M. and {Conte}, A. and {Conti}, L. and {Cook}, D. and {Corbitt}, T.~R. and {Cornish}, N. and {Corsi}, A. and {Cortese}, S. and {Costa}, C.~A. and {Coughlin}, M.~W. and {Coughlin}, S.~B. and {Coulon}, J. -P. and {Countryman}, S.~T. and {Couvares}, P. and {Cowan}, E.~E. and {Coward}, D.~M. and {Cowart}, M.~J. and {Coyne}, D.~C. and {Coyne}, R. and {Craig}, K. and {Creighton}, J.~D.~E. and {Creighton}, T.~D. and {Cripe}, J. and {Crowder}, S.~G. and {Cruise}, A.~M. and {Cumming}, A. and {Cunningham}, L. and {Cuoco}, E. and {Canton}, T. Dal and {Danilishin}, S.~L. and {D'Antonio}, S. and {Danzmann}, K. and {Darman}, N.~S. and {Da Silva Costa}, C.~F. and {Dattilo}, V. and {Dave}, I. and {Daveloza}, H.~P. and {Davier}, M. and {Davies}, G.~S. and {Daw}, E.~J. and {Day}, R. and {De}, S. and {DeBra}, D. and {Debreczeni}, G. and {Degallaix}, J. and {De Laurentis}, M. and {Del{\'e}glise}, S. and {Del Pozzo}, W. and {Denker}, T. and {Dent}, T. and {Dereli}, H. and {Dergachev}, V. and {DeRosa}, R.~T. and {De Rosa}, R. and {DeSalvo}, R. and {Dhurandhar}, S. and {D{\'\i}az}, M.~C. and {Di Fiore}, L. and {Di Giovanni}, M. and {Di Lieto}, A. and {Di Pace}, S. and {Di Palma}, I. and {Di Virgilio}, A. and {Dojcinoski}, G. and {Dolique}, V. and {Donovan}, F. and {Dooley}, K.~L. and {Doravari}, S. and {Douglas}, R. and {Downes}, T.~P. and {Drago}, M. and {Drever}, R.~W.~P. and {Driggers}, J.~C. and {Du}, Z. and {Ducrot}, M. and {Dwyer}, S.~E. and {Edo}, T.~B. and {Edwards}, M.~C. and {Effler}, A. and {Eggenstein}, H. -B. and {Ehrens}, P. and {Eichholz}, J. and {Eikenberry}, S.~S. and {Engels}, W. and {Essick}, R.~C. and {Etzel}, T. and {Evans}, M. and {Evans}, T.~M. and {Everett}, R. and {Factourovich}, M. and {Fafone}, V. and {Fair}, H. and {Fairhurst}, S. and {Fan}, X. and {Fang}, Q. and {Farinon}, S. and {Farr}, B. and {Farr}, W.~M. and {Favata}, M. and {Fays}, M. and {Fehrmann}, H. and {Fejer}, M.~M. and {Feldbaum}, D. and {Ferrante}, I. and {Ferreira}, E.~C. and {Ferrini}, F. and {Fidecaro}, F. and {Finn}, L.~S. and {Fiori}, I. and {Fiorucci}, D. and {Fisher}, R.~P. and {Flaminio}, R. and {Fletcher}, M. and {Fong}, H. and {Fournier}, J. -D. and {Franco}, S. and {Frasca}, S. and {Frasconi}, F. and {Frede}, M. and {Frei}, Z. and {Freise}, A. and {Frey}, R. and {Frey}, V. and {Fricke}, T.~T. and {Fritschel}, P. and {Frolov}, V.~V. and {Fulda}, P. and {Fyffe}, M. and {Gabbard}, H.~A.~G. and {Gair}, J.~R. and {Gammaitoni}, L. and {Gaonkar}, S.~G. and {Garufi}, F. and {Gatto}, A. and {Gaur}, G. and {Gehrels}, N. and {Gemme}, G. and {Gendre}, B. and {Genin}, E. and {Gennai}, A. and {George}, J. and {Gergely}, L. and {Germain}, V. and {Ghosh}, Abhirup and {Ghosh}, Archisman and {Ghosh}, S. and {Giaime}, J.~A. and {Giardina}, K.~D. and {Giazotto}, A. and {Gill}, K. and {Glaefke}, A. and {Gleason}, J.~R. and {Goetz}, E. and {Goetz}, R. and {Gondan}, L. and {Gonz{\'a}lez}, G. and {Gonzalez Castro}, J.~M. and {Gopakumar}, A. and {Gordon}, N.~A. and {Gorodetsky}, M.~L. and {Gossan}, S.~E. and {Gosselin}, M. and {Gouaty}, R. and {Graef}, C. and {Graff}, P.~B. and {Granata}, M. and {Grant}, A. and {Gras}, S. and {Gray}, C. and {Greco}, G. and {Green}, A.~C. and {Greenhalgh}, R.~J.~S. and {Groot}, P. and {Grote}, H. and {Grunewald}, S. and {Guidi}, G.~M. and {Guo}, X. and {Gupta}, A. and {Gupta}, M.~K. and {Gushwa}, K.~E. and {Gustafson}, E.~K. and {Gustafson}, R. and {Hacker}, J.~J. and {Hall}, B.~R. and {Hall}, E.~D. and {Hammond}, G. and {Haney}, M. and {Hanke}, M.~M. and {Hanks}, J. and {Hanna}, C. and {Hannam}, M.~D. and {Hanson}, J. and {Hardwick}, T. and {Harms}, J. and {Harry}, G.~M. and {Harry}, I.~W. and {Hart}, M.~J. and {Hartman}, M.~T. and {Haster}, C. -J. and {Haughian}, K. and {Healy}, J. and {Heefner}, J. and {Heidmann}, A. and {Heintze}, M.~C. and {Heinzel}, G. and {Heitmann}, H. and {Hello}, P. and {Hemming}, G. and {Hendry}, M. and {Heng}, I.~S. and {Hennig}, J. and {Heptonstall}, A.~W. and {Heurs}, M. and {Hild}, S. and {Hoak}, D. and {Hodge}, K.~A. and {Hofman}, D. and {Hollitt}, S.~E. and {Holt}, K. and {Holz}, D.~E. and {Hopkins}, P. and {Hosken}, D.~J. and {Hough}, J. and {Houston}, E.~A. and {Howell}, E.~J. and {Hu}, Y.~M. and {Huang}, S. and {Huerta}, E.~A. and {Huet}, D. and {Hughey}, B. and {Husa}, S. and {Huttner}, S.~H. and {Huynh-Dinh}, T. and {Idrisy}, A. and {Indik}, N. and {Ingram}, D.~R. and {Inta}, R. and {Isa}, H.~N. and {Isac}, J. -M. and {Isi}, M. and {Islas}, G. and {Isogai}, T. and {Iyer}, B.~R. and {Izumi}, K. and {Jacobson}, M.~B. and {Jacqmin}, T. and {Jang}, H. and {Jani}, K. and {Jaranowski}, P. and {Jawahar}, S. and {Jim{\'e}nez-Forteza}, F. and {Johnson}, W.~W. and {Johnson-McDaniel}, N.~K. and {Jones}, D.~I. and {Jones}, R. and {Jonker}, R.~J.~G. and {Ju}, L. and {Haris}, K. and {Kalaghatgi}, C.~V. and {Kalogera}, V. and {Kandhasamy}, S. and {Kang}, G. and {Kanner}, J.~B. and {Karki}, S. and {Kasprzack}, M. and {Katsavounidis}, E. and {Katzman}, W. and {Kaufer}, S. and {Kaur}, T. and {Kawabe}, K. and {Kawazoe}, F. and {K{\'e}f{\'e}lian}, F. and {Kehl}, M.~S. and {Keitel}, D. and {Kelley}, D.~B. and {Kells}, W. and {Kennedy}, R. and {Keppel}, D.~G. and {Key}, J.~S. and {Khalaidovski}, A. and {Khalili}, F.~Y. and {Khan}, I. and {Khan}, S. and {Khan}, Z. and {Khazanov}, E.~A. and {Kijbunchoo}, N. and {Kim}, C. and {Kim}, J. and {Kim}, K. and {Kim}, Nam-Gyu and {Kim}, Namjun and {Kim}, Y. -M. and {King}, E.~J. and {King}, P.~J. and {Kinzel}, D.~L. and {Kissel}, J.~S. and {Kleybolte}, L. and {Klimenko}, S. and {Koehlenbeck}, S.~M. and {Kokeyama}, K. and {Koley}, S. and {Kondrashov}, V. and {Kontos}, A. and {Koranda}, S. and {Korobko}, M. and {Korth}, W.~Z. and {Kowalska}, I. and {Kozak}, D.~B. and {Kringel}, V. and {Krishnan}, B. and {Kr{\'o}lak}, A. and {Krueger}, C. and {Kuehn}, G. and {Kumar}, P. and {Kumar}, R. and {Kuo}, L. and {Kutynia}, A. and {Kwee}, P. and {Lackey}, B.~D. and {Landry}, M. and {Lange}, J. and {Lantz}, B. and {Lasky}, P.~D. and {Lazzarini}, A. and {Lazzaro}, C. and {Leaci}, P. and {Leavey}, S. and {Lebigot}, E.~O. and {Lee}, C.~H. and {Lee}, H.~K. and {Lee}, H.~M. and {Lee}, K. and {Lenon}, A. and {Leonardi}, M. and {Leong}, J.~R. and {Leroy}, N. and {Letendre}, N. and {Levin}, Y. and {Levine}, B.~M. and {Li}, T.~G.~F. and {Libson}, A. and {Littenberg}, T.~B. and {Lockerbie}, N.~A. and {Logue}, J. and {Lombardi}, A.~L. and {London}, L.~T. and {Lord}, J.~E. and {Lorenzini}, M. and {Loriette}, V. and {Lormand}, M. and {Losurdo}, G. and {Lough}, J.~D. and {Lousto}, C.~O. and {Lovelace}, G. and {L{\"u}ck}, H. and {Lundgren}, A.~P. and {Luo}, J. and {Lynch}, R. and {Ma}, Y. and {MacDonald}, T. and {Machenschalk}, B. and {MacInnis}, M. and {Macleod}, D.~M. and {Maga{\~n}a-Sandoval}, F. and {Magee}, R.~M. and {Mageswaran}, M. and {Majorana}, E. and {Maksimovic}, I. and {Malvezzi}, V. and {Man}, N. and {Mandel}, I. and {Mandic}, V. and {Mangano}, V. and {Mansell}, G.~L. and {Manske}, M. and {Mantovani}, M. and {Marchesoni}, F. and {Marion}, F. and {M{\'a}rka}, S. and {M{\'a}rka}, Z. and {Markosyan}, A.~S. and {Maros}, E. and {Martelli}, F. and {Martellini}, L. and {Martin}, I.~W. and {Martin}, R.~M. and {Martynov}, D.~V. and {Marx}, J.~N. and {Mason}, K. and {Masserot}, A. and {Massinger}, T.~J. and {Masso-Reid}, M. and {Matichard}, F. and {Matone}, L. and {Mavalvala}, N. and {Mazumder}, N. and {Mazzolo}, G. and {McCarthy}, R. and {McClelland}, D.~E. and {McCormick}, S. and {McGuire}, S.~C. and {McIntyre}, G. and {McIver}, J. and {McManus}, D.~J. and {McWilliams}, S.~T. and {Meacher}, D. and {Meadors}, G.~D. and {Meidam}, J. and {Melatos}, A. and {Mendell}, G. and {Mendoza-Gandara}, D. and {Mercer}, R.~A. and {Merilh}, E. and {Merzougui}, M. and {Meshkov}, S. and {Messenger}, C. and {Messick}, C. and {Meyers}, P.~M. and {Mezzani}, F. and {Miao}, H. and {Michel}, C. and {Middleton}, H. and {Mikhailov}, E.~E. and {Milano}, L. and {Miller}, J. and {Millhouse}, M. and {Minenkov}, Y. and {Ming}, J. and {Mirshekari}, S. and {Mishra}, C. and {Mitra}, S. and {Mitrofanov}, V.~P. and {Mitselmakher}, G. and {Mittleman}, R. and {Moggi}, A. and {Mohan}, M. and {Mohapatra}, S.~R.~P. and {Montani}, M. and {Moore}, B.~C. and {Moore}, C.~J. and {Moraru}, D. and {Moreno}, G. and {Morriss}, S.~R. and {Mossavi}, K. and {Mours}, B. and {Mow-Lowry}, C.~M. and {Mueller}, C.~L. and {Mueller}, G. and {Muir}, A.~W. and {Mukherjee}, Arunava and {Mukherjee}, D. and {Mukherjee}, S. and {Mukund}, N. and {Mullavey}, A. and {Munch}, J. and {Murphy}, D.~J. and {Murray}, P.~G. and {Mytidis}, A. and {Nardecchia}, I. and {Naticchioni}, L. and {Nayak}, R.~K. and {Necula}, V. and {Nedkova}, K. and {Nelemans}, G. and {Neri}, M. and {Neunzert}, A. and {Newton}, G. and {Nguyen}, T.~T. and {Nielsen}, A.~B. and {Nissanke}, S. and {Nitz}, A. and {Nocera}, F. and {Nolting}, D. and {Normandin}, M.~E.~N. and {Nuttall}, L.~K. and {Oberling}, J. and {Ochsner}, E. and {O'Dell}, J. and {Oelker}, E. and {Ogin}, G.~H. and {Oh}, J.~J. and {Oh}, S.~H. and {Ohme}, F. and {Oliver}, M. and {Oppermann}, P. and {Oram}, Richard J. and {O'Reilly}, B. and {O'Shaughnessy}, R. and {Ott}, C.~D. and {Ottaway}, D.~J. and {Ottens}, R.~S. and {Overmier}, H. and {Owen}, B.~J. and {Pai}, A. and {Pai}, S.~A. and {Palamos}, J.~R. and {Palashov}, O. and {Palomba}, C. and {Pal-Singh}, A. and {Pan}, H. and {Pan}, Y. and {Pankow}, C. and {Pannarale}, F. and {Pant}, B.~C. and {Paoletti}, F. and {Paoli}, A. and {Papa}, M.~A. and {Paris}, H.~R. and {Parker}, W. and {Pascucci}, D. and {Pasqualetti}, A. and {Passaquieti}, R. and {Passuello}, D. and {Patricelli}, B. and {Patrick}, Z. and {Pearlstone}, B.~L. and {Pedraza}, M. and {Pedurand}, R. and {Pekowsky}, L. and {Pele}, A. and {Penn}, S. and {Perreca}, A. and {Pfeiffer}, H.~P. and {Phelps}, M. and {Piccinni}, O. and {Pichot}, M. and {Pickenpack}, M. and {Piergiovanni}, F. and {Pierro}, V. and {Pillant}, G. and {Pinard}, L. and {Pinto}, I.~M. and {Pitkin}, M. and {Poeld}, J.~H. and {Poggiani}, R. and {Popolizio}, P. and {Post}, A. and {Powell}, J. and {Prasad}, J. and {Predoi}, V. and {Premachandra}, S.~S. and {Prestegard}, T. and {Price}, L.~R. and {Prijatelj}, M. and {Principe}, M. and {Privitera}, S. and {Prix}, R. and {Prodi}, G.~A. and {Prokhorov}, L. and {Puncken}, O. and {Punturo}, M. and {Puppo}, P. and {P{\"u}rrer}, M. and {Qi}, H. and {Qin}, J. and {Quetschke}, V. and {Quintero}, E.~A. and {Quitzow-James}, R. and {Raab}, F.~J. and {Rabeling}, D.~S. and {Radkins}, H. and {Raffai}, P. and {Raja}, S. and {Rakhmanov}, M. and {Ramet}, C.~R. and {Rapagnani}, P. and {Raymond}, V. and {Razzano}, M. and {Re}, V. and {Read}, J. and {Reed}, C.~M. and {Regimbau}, T. and {Rei}, L. and {Reid}, S. and {Reitze}, D.~H. and {Rew}, H. and {Reyes}, S.~D. and {Ricci}, F. and {Riles}, K. and {Robertson}, N.~A. and {Robie}, R. and {Robinet}, F. and {Rocchi}, A. and {Rolland}, L. and {Rollins}, J.~G. and {Roma}, V.~J. and {Romano}, J.~D. and {Romano}, R. and {Romanov}, G. and {Romie}, J.~H. and {Rosi{\'n}ska}, D. and {Rowan}, S. and {R{\"u}diger}, A. and {Ruggi}, P. and {Ryan}, K. and {Sachdev}, S. and {Sadecki}, T. and {Sadeghian}, L. and {Salconi}, L. and {Saleem}, M. and {Salemi}, F. and {Samajdar}, A. and {Sammut}, L. and {Sampson}, L.~M. and {Sanchez}, E.~J. and {Sandberg}, V. and {Sandeen}, B. and {Sanders}, G.~H. and {Sanders}, J.~R. and {Sassolas}, B. and {Sathyaprakash}, B.~S. and {Saulson}, P.~R. and {Sauter}, O. and {Savage}, R.~L. and {Sawadsky}, A. and {Schale}, P. and {Schilling}, R. and {Schmidt}, J. and {Schmidt}, P. and {Schnabel}, R. and {Schofield}, R.~M.~S. and {Sch{\"o}nbeck}, A. and {Schreiber}, E. and {Schuette}, D. and {Schutz}, B.~F. and {Scott}, J. and {Scott}, S.~M. and {Sellers}, D. and {Sengupta}, A.~S. and {Sentenac}, D. and {Sequino}, V. and {Sergeev}, A. and {Serna}, G. and {Setyawati}, Y. and {Sevigny}, A. and {Shaddock}, D.~A. and {Shaffer}, T. and {Shah}, S. and {Shahriar}, M.~S. and {Shaltev}, M. and {Shao}, Z. and {Shapiro}, B. and {Shawhan}, P. and {Sheperd}, A. and {Shoemaker}, D.~H. and {Shoemaker}, D.~M. and {Siellez}, K. and {Siemens}, X. and {Sigg}, D. and {Silva}, A.~D. and {Simakov}, D. and {Singer}, A. and {Singer}, L.~P. and {Singh}, A. and {Singh}, R. and {Singhal}, A. and {Sintes}, A.~M. and {Slagmolen}, B.~J.~J. and {Smith}, J.~R. and {Smith}, M.~R. and {Smith}, N.~D. and {Smith}, R.~J.~E. and {Son}, E.~J. and {Sorazu}, B. and {Sorrentino}, F. and {Souradeep}, T. and {Srivastava}, A.~K. and {Staley}, A. and {Steinke}, M. and {Steinlechner}, J. and {Steinlechner}, S. and {Steinmeyer}, D. and {Stephens}, B.~C. and {Stevenson}, S.~P. and {Stone}, R. and {Strain}, K.~A. and {Straniero}, N. and {Stratta}, G. and {Strauss}, N.~A. and {Strigin}, S. and {Sturani}, R. and {Stuver}, A.~L. and {Summerscales}, T.~Z. and {Sun}, L. and {Sutton}, P.~J. and {Swinkels}, B.~L. and {Szczepa{\'n}czyk}, M.~J. and {Tacca}, M. and {Talukder}, D. and {Tanner}, D.~B. and {T{\'a}pai}, M. and {Tarabrin}, S.~P. and {Taracchini}, A. and {Taylor}, R. and {Theeg}, T. and {Thirugnanasambandam}, M.~P. and {Thomas}, E.~G. and {Thomas}, M. and {Thomas}, P. and {Thorne}, K.~A. and {Thorne}, K.~S. and {Thrane}, E. and {Tiwari}, S. and {Tiwari}, V. and {Tokmakov}, K.~V. and {Tomlinson}, C. and {Tonelli}, M. and {Torres}, C.~V. and {Torrie}, C.~I. and {T{\"o}yr{\"a}}, D. and {Travasso}, F. and {Traylor}, G. and {Trifir{\`o}}, D. and {Tringali}, M.~C. and {Trozzo}, L. and {Tse}, M. and {Turconi}, M. and {Tuyenbayev}, D. and {Ugolini}, D. and {Unnikrishnan}, C.~S. and {Urban}, A.~L. and {Usman}, S.~A. and {Vahlbruch}, H. and {Vajente}, G. and {Valdes}, G. and {Vallisneri}, M. and {van Bakel}, N. and {van Beuzekom}, M. and {van den Brand}, J.~F.~J. and {Van Den Broeck}, C. and {Vander-Hyde}, D.~C. and {van der Schaaf}, L. and {van Heijningen}, J.~V. and {van Veggel}, A.~A. and {Vardaro}, M. and {Vass}, S. and {Vas{\'u}th}, M. and {Vaulin}, R. and {Vecchio}, A. and {Vedovato}, G. and {Veitch}, J. and {Veitch}, P.~J. and {Venkateswara}, K. and {Verkindt}, D. and {Vetrano}, F. and {Vicer{\'e}}, A. and {Vinciguerra}, S. and {Vine}, D.~J. and {Vinet}, J. -Y. and {Vitale}, S. and {Vo}, T. and {Vocca}, H. and {Vorvick}, C. and {Voss}, D. and {Vousden}, W.~D. and {Vyatchanin}, S.~P. and {Wade}, A.~R. and {Wade}, L.~E. and {Wade}, M. and {Waldman}, S.~J. and {Walker}, M. and {Wallace}, L. and {Walsh}, S. and {Wang}, G. and {Wang}, H. and {Wang}, M. and {Wang}, X. and {Wang}, Y. and {Ward}, H. and {Ward}, R.~L. and {Warner}, J. and {Was}, M. and {Weaver}, B. and {Wei}, L. -W. and {Weinert}, M. and {Weinstein}, A.~J. and {Weiss}, R. and {Welborn}, T. and {Wen}, L. and {We{\ss}els}, P. and {Westphal}, T. and {Wette}, K. and {Whelan}, J.~T. and {Whitcomb}, S.~E. and {White}, D.~J. and {Whiting}, B.~F. and {Wiesner}, K. and {Wilkinson}, C. and {Willems}, P.~A. and {Williams}, L. and {Williams}, R.~D. and {Williamson}, A.~R. and {Willis}, J.~L. and {Willke}, B. and {Wimmer}, M.~H. and {Winkelmann}, L. and {Winkler}, W. and {Wipf}, C.~C. and {Wiseman}, A.~G. and {Wittel}, H. and {Woan}, G. and {Worden}, J. and {Wright}, J.~L. and {Wu}, G. and {Yablon}, J. and {Yakushin}, I. and {Yam}, W. and {Yamamoto}, H. and {Yancey}, C.~C. and {Yap}, M.~J. and {Yu}, H. and {Yvert}, M. and {Zadro{\.Z}ny}, A. and {Zangrando}, L. and {Zanolin}, M. and {Zendri}, J. -P. and {Zevin}, M. and {Zhang}, F. and {Zhang}, L. and {Zhang}, M. and {Zhang}, Y. and {Zhao}, C. and {Zhou}, M. and {Zhou}, Z. and {Zhu}, X.~J. and {Zucker}, M.~E. and {Zuraw}, S.~E. and {Zweizig}, J. and {LIGO Scientific Collaboration} and {Virgo Collaboration}},
        title = "{Observation of Gravitational Waves from a Binary Black Hole Merger}",
      journal = {\prl},
         year = 2016,
        month = feb,
       volume = {116},
       number = {6},
          eid = {061102},
        pages = {061102},
          doi = {10.1103/PhysRevLett.116.061102},
archivePrefix = {arXiv},
       eprint = {1602.03837},
 primaryClass = {gr-qc},
       adsurl = {https://ui.adsabs.harvard.edu/abs/2016PhRvL.116f1102A}
}

@ARTICLE{2008IAUC.8972....1N,
       author = {{Nakano}, S. and {Nishiyama}, K. and {Kabashima}, F. and {Sakurai}, Y. and {Jacques}, C. and {Pimentel}, E. and {Chekhovich}, D. and {Korotkiy}, S. and {Kryachko}, T. and {Samus}, N.~N.},
        title = "{V1309 Scorpii = Nova Scorpii 2008}",
      journal = {\iaucirc},
         year = 2008,
        month = sep,
       volume = {8972},
        pages = {1},
       adsurl = {https://ui.adsabs.harvard.edu/abs/2008IAUC.8972....1N}
}

@ARTICLE{2011A&A...528A.114T,
       author = {{Tylenda}, R. and {Hajduk}, M. and {Kami{\'n}ski}, T. and {Udalski}, A. and {Soszy{\'n}ski}, I. and {Szyma{\'n}ski}, M.~K. and {Kubiak}, M. and {Pietrzy{\'n}ski}, G. and {Poleski}, R. and {Wyrzykowski}, {\L}. and {Ulaczyk}, K.},
        title = "{V1309 Scorpii: merger of a contact binary}",
      journal = {\aap},
         year = 2011,
        month = apr,
       volume = {528},
          eid = {A114},
        pages = {A114},
          doi = {10.1051/0004-6361/201016221},
archivePrefix = {arXiv},
       eprint = {1012.0163},
 primaryClass = {astro-ph.SR},
       adsurl = {https://ui.adsabs.harvard.edu/abs/2011A&A...528A.114T}
}

@ARTICLE{2006AJ....131..621G,
       author = {{Gettel}, S.~J. and {Geske}, M.~T. and {McKay}, T.~A.},
        title = "{A Catalog of 1022 Bright Contact Binary Stars}",
      journal = {\aj},
         year = 2006,
        month = jan,
       volume = {131},
       number = {1},
        pages = {621-632},
          doi = {10.1086/498016},
archivePrefix = {arXiv},
       eprint = {astro-ph/0509819},
 primaryClass = {astro-ph},
       adsurl = {https://ui.adsabs.harvard.edu/abs/2006AJ....131..621G}
}

@ARTICLE{2020ApJS..249...18C,
       author = {{Chen}, Xiaodian and {Wang}, Shu and {Deng}, Licai and {de Grijs}, Richard and {Yang}, Ming and {Tian}, Hao},
        title = "{The Zwicky Transient Facility Catalog of Periodic Variable Stars}",
      journal = {\apjs},
         year = 2020,
        month = jul,
       volume = {249},
       number = {1},
          eid = {18},
        pages = {18},
          doi = {10.3847/1538-4365/ab9cae},
archivePrefix = {arXiv},
       eprint = {2005.08662},
 primaryClass = {astro-ph.SR},
       adsurl = {https://ui.adsabs.harvard.edu/abs/2020ApJS..249...18C}
}

@ARTICLE{2020RAA....20..163Q,
       author = {{Qian}, Sheng-Bang and {Zhu}, Li-Ying and {Liu}, Liang and {Zhang}, Xu-Dong and {Shi}, Xiang-Dong and {He}, Jia-Jia and {Zhang}, Jia},
        title = "{Contact binaries at different evolutionary stages}",
      journal = {Research in Astronomy and Astrophysics},
         year = 2020,
        month = oct,
       volume = {20},
       number = {10},
          eid = {163},
        pages = {163},
          doi = {10.1088/1674-4527/20/10/163},
       adsurl = {https://ui.adsabs.harvard.edu/abs/2020RAA....20..163Q}
}

@ARTICLE{2021MNRAS.503.3975P,
       author = {{Petrosky}, Evan and {Hwang}, Hsiang-Chih and {Zakamska}, Nadia L. and {Chandra}, Vedant and {Hill}, Matthew J.},
        title = "{Variability, periodicity, and contact binaries in WISE}",
      journal = {\mnras},
         year = 2021,
        month = may,
       volume = {503},
       number = {3},
        pages = {3975-3991},
          doi = {10.1093/mnras/stab592},
archivePrefix = {arXiv},
       eprint = {2012.04690},
 primaryClass = {astro-ph.SR},
       adsurl = {https://ui.adsabs.harvard.edu/abs/2021MNRAS.503.3975P}
}

@INPROCEEDINGS{2014SPIE.9143E..20R,
       author = {{Ricker}, George R. and {Winn}, Joshua N. and {Vanderspek}, Roland and {Latham}, David W. and {Bakos}, G{\'a}sp{\'a}r. {\'A}. and {Bean}, Jacob L. and {Berta-Thompson}, Zachory K. and {Brown}, Timothy M. and {Buchhave}, Lars and {Butler}, Nathaniel R. and {Butler}, R. Paul and {Chaplin}, William J. and {Charbonneau}, David and {Christensen-Dalsgaard}, J{\o}rgen and {Clampin}, Mark and {Deming}, Drake and {Doty}, John and {De Lee}, Nathan and {Dressing}, Courtney and {Dunham}, E.~W. and {Endl}, Michael and {Fressin}, Francois and {Ge}, Jian and {Henning}, Thomas and {Holman}, Matthew J. and {Howard}, Andrew W. and {Ida}, Shigeru and {Jenkins}, Jon and {Jernigan}, Garrett and {Johnson}, John A. and {Kaltenegger}, Lisa and {Kawai}, Nobuyuki and {Kjeldsen}, Hans and {Laughlin}, Gregory and {Levine}, Alan M. and {Lin}, Douglas and {Lissauer}, Jack J. and {MacQueen}, Phillip and {Marcy}, Geoffrey and {McCullough}, P.~R. and {Morton}, Timothy D. and {Narita}, Norio and {Paegert}, Martin and {Palle}, Enric and {Pepe}, Francesco and {Pepper}, Joshua and {Quirrenbach}, Andreas and {Rinehart}, S.~A. and {Sasselov}, Dimitar and {Sato}, Bun'ei and {Seager}, Sara and {Sozzetti}, Alessandro and {Stassun}, Keivan G. and {Sullivan}, Peter and {Szentgyorgyi}, Andrew and {Torres}, Guillermo and {Udry}, Stephane and {Villasenor}, Joel},
        title = "{Transiting Exoplanet Survey Satellite (TESS)}",
    booktitle = {Space Telescopes and Instrumentation 2014: Optical, Infrared, and Millimeter Wave},
         year = 2014,
       editor = {{Oschmann}, Jacobus M., Jr. and {Clampin}, Mark and {Fazio}, Giovanni G. and {MacEwen}, Howard A.},
       series = {Society of Photo-Optical Instrumentation Engineers (SPIE) Conference Series},
       volume = {9143},
        month = aug,
          eid = {914320},
        pages = {914320},
          doi = {10.1117/12.2063489},
archivePrefix = {arXiv},
       eprint = {1406.0151},
 primaryClass = {astro-ph.EP},
       adsurl = {https://ui.adsabs.harvard.edu/abs/2014SPIE.9143E..20R}
}

@INPROCEEDINGS{2016SPIE.9913E..3EJ,
       author = {{Jenkins}, Jon M. and {Twicken}, Joseph D. and {McCauliff}, Sean and {Campbell}, Jennifer and {Sanderfer}, Dwight and {Lung}, David and {Mansouri-Samani}, Masoud and {Girouard}, Forrest and {Tenenbaum}, Peter and {Klaus}, Todd and {Smith}, Jeffrey C. and {Caldwell}, Douglas A. and {Chacon}, A.~D. and {Henze}, Christopher and {Heiges}, Cory and {Latham}, David W. and {Morgan}, Edward and {Swade}, Daryl and {Rinehart}, Stephen and {Vanderspek}, Roland},
        title = "{The TESS science processing operations center}",
    booktitle = {Software and Cyberinfrastructure for Astronomy IV},
         year = 2016,
       editor = {{Chiozzi}, Gianluca and {Guzman}, Juan C.},
       series = {Society of Photo-Optical Instrumentation Engineers (SPIE) Conference Series},
       volume = {9913},
        month = aug,
          eid = {99133E},
        pages = {99133E},
          doi = {10.1117/12.2233418},
       adsurl = {https://ui.adsabs.harvard.edu/abs/2016SPIE.9913E..3EJ}
}
\bibliographystyle{aasjournal}

\end{document}